\pdfoutput=1
\documentclass{article}

\PassOptionsToPackage{numbers, square}{natbib}
\usepackage[final]{neurips_2022}

\usepackage[utf8]{inputenc} 
\usepackage[T1]{fontenc}    
\usepackage{hyperref}       
\usepackage{url}            
\usepackage{booktabs}       
\usepackage{amsfonts}       
\usepackage{nicefrac}       
\usepackage{microtype}      
\usepackage[dvipsnames]{xcolor}         
\hypersetup{colorlinks=true,linkcolor=red,anchorcolor=black,citecolor=NavyBlue,filecolor=black,menucolor=black,runcolor=black,urlcolor=black}

\usepackage{graphicx}
\usepackage{bmpsize}
\usepackage{subcaption}
\usepackage{amsmath}
\usepackage{amssymb}
\usepackage{mathtools}
\usepackage{amsthm}
\usepackage{color}
\usepackage{diagbox}
\usepackage{thm-restate}
\usepackage{multirow}
\usepackage{forloop}
\usepackage{cleveref}

\theoremstyle{plain}

\theoremstyle{definition}

\theoremstyle{remark}

\usepackage{amsmath,amsfonts,bm,amssymb,mathtools,amsthm}
\usepackage{color,xcolor,xspace}
\usepackage{booktabs}
\usepackage{thm-restate}
\usepackage{bbding}
\usepackage{pifont}
\usepackage{wasysym}
\usepackage[most]{tcolorbox}

\newcommand{\bb}[1]{{\mathbb{#1}}}
\newcommand{\norm}[1]{{\lVert {#1} \rVert}}

\def\eqref#1{Eq.(\ref{#1})}

\def\1{\bm{1}}

\def\rvx{{\mathbf{x}}}
\def\rvy{{\mathbf{y}}}
\def\rvz{{\mathbf{z}}}

\def\vb{{\bm{b}}}
\def\vc{{\bm{c}}}

\def\vk{{\bm{k}}}

\def\vp{{\bm{p}}}

\def\vr{{\bm{r}}}

\def\vx{{\bm{x}}}

\def\mA{{\bm{A}}}
\def\mB{{\bm{B}}}

\def\mH{{\bm{H}}}
\def\mI{{\bm{I}}}

\def\mK{{\bm{K}}}
\def\mL{{\bm{L}}}
\def\mM{{\bm{M}}}
\def\mN{{\bm{N}}}

\def\mP{{\bm{P}}}

\def\mU{{\bm{U}}}
\def\mV{{\bm{V}}}

\def\mX{{\bm{X}}}

\def\mSigma{{\bm{\Sigma}}}

\DeclareMathAlphabet{\mathsfit}{\encodingdefault}{\sfdefault}{m}{sl}
\SetMathAlphabet{\mathsfit}{bold}{\encodingdefault}{\sfdefault}{bx}{n}

\def\gN{{\mathcal{N}}}

\def\sR{{\mathbb{R}}}

\newcommand{\KL}{D_{\mathrm{KL}}}

\definecolor{first}{rgb}{0.1, 0.42, 0.1}
\definecolor{second}{rgb}{0.0, 0.35, 0.56}

\newcommand{\first}[1]{{\color{first}$\mathbf{#1}$}}
\newcommand{\second}[1]{{\color{second}${#1}$}}

\title{Denoising Diffusion Restoration Models}

\author{%
  Bahjat Kawar \\
  Department of Computer Science\\
  Technion, Haifa, Israel\\
  \texttt{bahjat.kawar@cs.technion.ac.il} \\
  \And
  Michael Elad \\
  Department of Computer Science\\
  Technion, Haifa, Israel\\
  \texttt{elad@cs.technion.ac.il} \\
  \And
  Stefano Ermon \\
  Department of Computer Science\\
  Stanford, California, USA\\
  \texttt{ermon@cs.stanford.edu} \\
  \And
  Jiaming Song \\
  NVIDIA\\
  Santa Clara, California, USA\\
  \texttt{jiamings@nvidia.com} \\
}

\begin{document}

\maketitle

\begin{abstract}
Many interesting tasks in image restoration can be cast as linear inverse problems.
A recent family of approaches for solving these problems uses stochastic algorithms that sample from the posterior distribution of natural images given the measurements.
However, efficient solutions often require problem-specific supervised training to model the posterior, whereas unsupervised methods that are not problem-specific typically rely on inefficient iterative methods.
This work addresses these issues by introducing Denoising Diffusion Restoration Models (DDRM), an efficient, unsupervised posterior sampling method.
Motivated by variational inference, DDRM takes advantage of a pre-trained denoising diffusion generative model for solving any linear inverse problem.
We demonstrate DDRM's versatility on several image datasets for super-resolution, deblurring, inpainting, and colorization under various amounts of measurement noise.
DDRM outperforms the current leading unsupervised methods on the diverse ImageNet dataset in reconstruction quality, perceptual quality, and runtime, being $5\times$ faster than the nearest competitor.
DDRM also generalizes well for natural images out of the distribution of the observed ImageNet training set.\footnote{Project website: \url{https://ddrm-ml.github.io/}}
\end{abstract}

\newcounter{row}
\newcounter{col}

\section{Introduction}

Many problems in image processing, including super-resolution~\cite{ledig2017photo,haris2018deep}, deblurring~\cite{kupyn2019deblurgan,suin2020spatially}, inpainting~\cite{yeh2017semantic}, colorization~\cite{larsson2016learning,zhang2016colorful}, and compressive sensing~\cite{baraniuk2007compressive}, are instances of 
linear inverse problems, where the goal is to recover an image from potentially noisy measurements given through a known linear degradation model.
For a specific degradation model, image restoration can be addressed through end-to-end \textit{supervised} training of neural networks, using pairs of original and degraded images~\cite{dong2015image,zhang2016colorful,palette}. However, real-world applications such as medical imaging often require flexibility to cope with multiple, possibly infinite, degradation models~\cite{song2021medical}. Here, \textit{unsupervised} approaches based on learned priors~\cite{ongie2020deep}, where the degradation model is only known and used during inference, may be more desirable since they can adapt to the given problem without re-training~\cite{pnp}.
By learning sound assumptions over the underlying structure of images (\textit{e.g.}, priors, proximal operators or denoisers), unsupervised approaches can achieve effective restoration without training on specific degradation models~\cite{pnp,red}.

Under this unsupervised 
setting, priors based on deep neural networks have demonstrated impressive empirical results in various image restoration tasks~\cite{red,ulyanov2018deep,santurkar2019image,dgp,gu2020image}.
To recover the signal, most existing methods obtain a prior-related term over the signal from a neural network (\textit{e.g.}, the distribution of natural images), and a likelihood term from the degradation model. They combine the two terms to form a posterior over the signal, and the inverse problem can be posed as solving an optimization problem (\textit{e.g.}, maximum a posteriori~\cite{calvetti2008hypermodels,red}) or solving a sampling problem (\textit{e.g.}, posterior sampling~\cite{bardsley2012mcmc,bardsley2014randomize,snips}).
Then, these problems are often solved with iterative methods, such as gradient descent or Langevin dynamics, which may be demanding in computation and sensitive to hyperparameter tuning. An extreme example is found in \cite{laumont} where a ``fast'' version of the algorithm uses $15,000$ neural function evaluations (NFEs). 

\newcommand{\imwidth}[0]{1.3cm}
\begin{figure*}
    \centering
    \def\arraystretch{1.1}
    \setlength\tabcolsep{0.03cm}
    \begin{subfigure}{0.48\textwidth}
    \centering
    \begin{tcolorbox}[size=small,height=6.65cm,valign=center]
    \begin{tabular}{lcccc}
        
        \raisebox{0.85cm}[0pt][0pt]{\rotatebox[origin=c]{90}{$4\times$}}
        & \includegraphics[width=\imwidth,height=\imwidth]{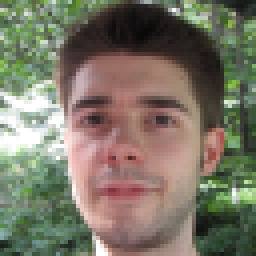}
        & \includegraphics[width=\imwidth,height=\imwidth]{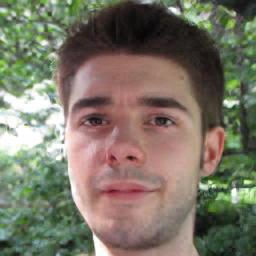} \hspace{0.15cm}
        & \includegraphics[width=\imwidth,height=\imwidth]{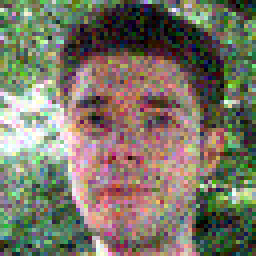}
        & \includegraphics[width=\imwidth,height=\imwidth]{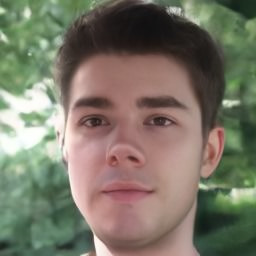}
        \\
        
        \raisebox{0.85cm}[0pt][0pt]{\rotatebox[origin=c]{90}{$8\times$}}
        & \includegraphics[width=\imwidth,height=\imwidth]{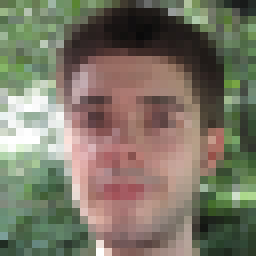}
        & \includegraphics[width=\imwidth,height=\imwidth]{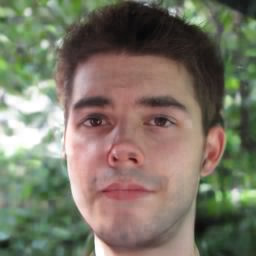} \hspace{0.15cm}
        & \includegraphics[width=\imwidth,height=\imwidth]{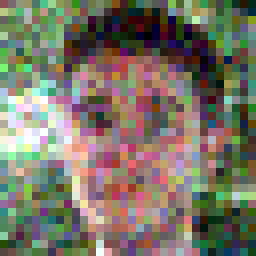}
        & \includegraphics[width=\imwidth,height=\imwidth]{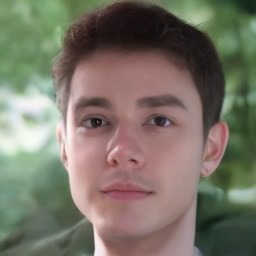}
        \\
        
        \raisebox{0.85cm}[0pt][0pt]{\rotatebox[origin=c]{90}{$16\times$}}
        & \includegraphics[width=\imwidth,height=\imwidth]{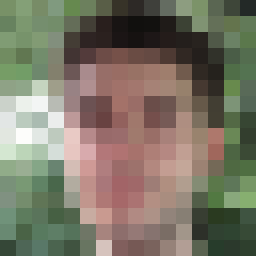}
        & \includegraphics[width=\imwidth,height=\imwidth]{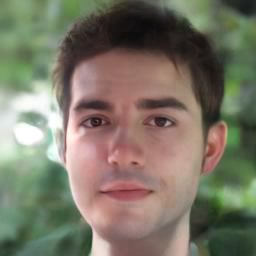} \hspace{0.15cm}
        & \includegraphics[width=\imwidth,height=\imwidth]{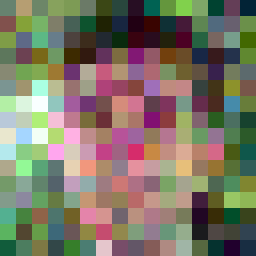}
        & \includegraphics[width=\imwidth,height=\imwidth]{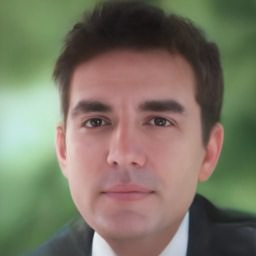}
        \\
        
        & \multicolumn{2}{c}{Noiseless} & \multicolumn{2}{c}{Noisy with $\sigma_{\rvy} = 0.1$}
    \end{tabular}
    \caption{Super-resolution}
    \label{fig:demo-sr}
    \end{tcolorbox}
    \end{subfigure}
    ~
    \begin{subfigure}{0.48\linewidth}
    \begin{tcolorbox}[size=small,valign=center]
    \centering
    \begin{tabular}{cccc}
         \includegraphics[width=\imwidth,height=\imwidth]{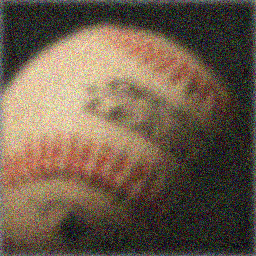}
        & \includegraphics[width=\imwidth,height=\imwidth]{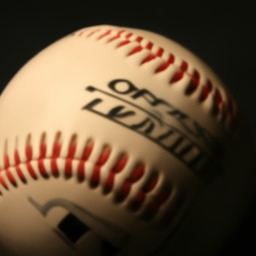} \hspace{0.2cm}
        & \includegraphics[width=\imwidth,height=\imwidth]{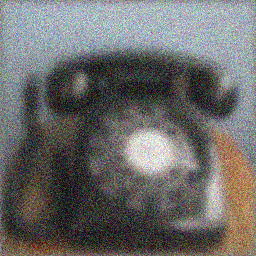}
        & \includegraphics[width=\imwidth,height=\imwidth]{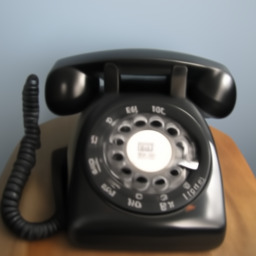}
    \end{tabular}
    \caption{Deblurring (Noisy with $\sigma_{\rvy} = 0.1$)}
    \end{tcolorbox}
    \begin{tcolorbox}[size=small,valign=center]
    \centering
    \begin{tabular}{cccc}
         \includegraphics[width=\imwidth,height=\imwidth]{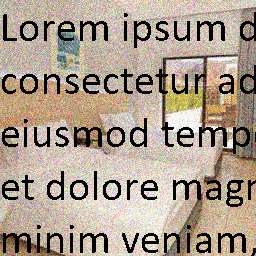}
        & \includegraphics[width=\imwidth,height=\imwidth]{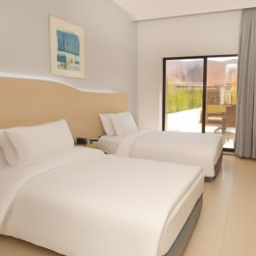} \hspace{0.2cm}
        & \includegraphics[width=\imwidth,height=\imwidth]{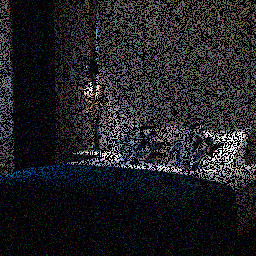}
        & \includegraphics[width=\imwidth,height=\imwidth]{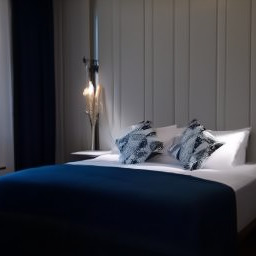}
    \end{tabular}
    \caption{Inpainting (Noisy with $\sigma_{\rvy} = 0.1$)}
    \end{tcolorbox}
    \begin{tcolorbox}[size=small,valign=center]
    \centering
    \begin{tabular}{cccc}
         \includegraphics[width=\imwidth,height=\imwidth]{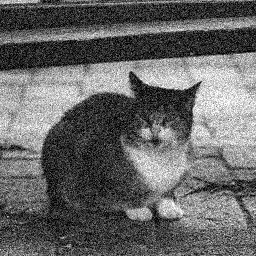}
        & \includegraphics[width=\imwidth,height=\imwidth]{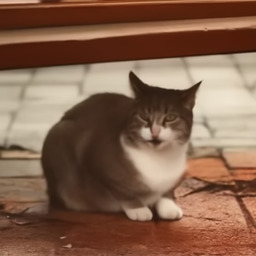} \hspace{0.2cm}
        & \includegraphics[width=\imwidth,height=\imwidth]{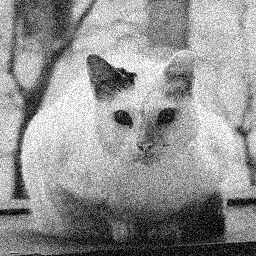}
        & \includegraphics[width=\imwidth,height=\imwidth]{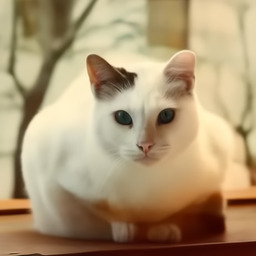}
    \end{tabular}
    \caption{Colorization (Noisy with $\sigma_{\rvy} = 0.1$)}
    \end{tcolorbox}
    \end{subfigure}
    \caption{Pairs of measurements and recovered images with a 20-step DDRM on super-resolution, deblurring, inpainting, and colorization, with or without noise, and with unconditional generative models. The images are not accessed during training.
    }
    \label{fig:demo}
\end{figure*}

\begin{figure*}
\centering
\includegraphics[width=0.95\textwidth]{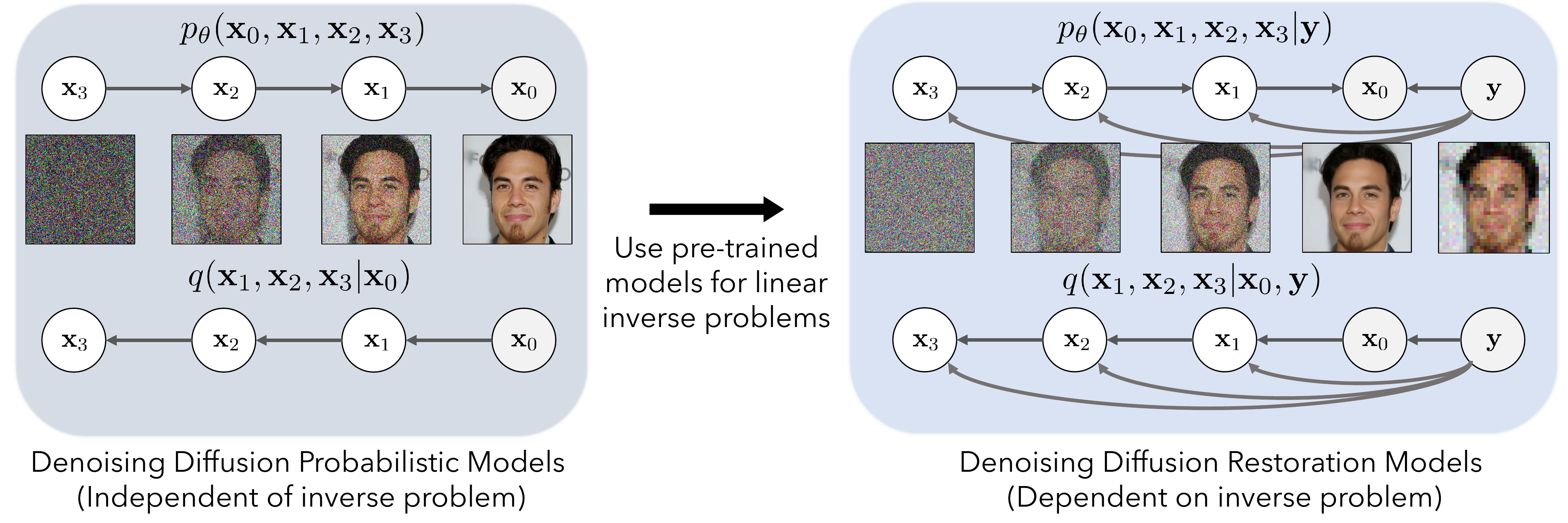}
\vspace{-0.2em}
\caption{Illustration of our DDRM method for a specific inverse problem (super-resolution + denoising). We can use unsupervised DDPM models as a good solution to the DDRM objective.
}
\label{fig:ddrm-figures}
\end{figure*}

Inspired by this unsupervised line of work, we introduce an efficient approach named Denoising Diffusion Restoration Models (DDRM), that can achieve competitive results in as low as $20$ NFEs. DDRM is a denoising diffusion generative model~\cite{sohl-dickstein2015deep,ddpm,song2020denoising} that gradually and stochastically denoises a sample to the desired output, conditioned on the measurements and the inverse problem.
This way we introduce a variational inference objective for learning the posterior distribution of the inverse problem at hand.
We then show its equivalence to the objective of an unconditional denoising diffusion generative model~\cite{ddpm}, which enables us to deploy such models in DDRM for various linear inverse problems (see \Cref{fig:ddrm-figures}). To our best knowledge, DDRM is the first general sampling-based inverse problem solver that can efficiently produce a range of high-quality, diverse, yet valid solutions for general content images.

We demonstrate the empirical effectiveness of DDRM by comparing with various competitive methods based on learned priors, such as Deep Generative Prior (DGP)~\cite{dgp}, SNIPS~\cite{snips}, and Regularization by Denoising (RED) \cite{red}.
On ImageNet examples, DDRM mostly outperforms the  neural network baselines under noiseless super-resolution and deblurring measured in PSNR and KID~\cite{kid}, and is at least $50\times$ more efficient in terms of NFEs when it is second-best.
Our advantage becomes even larger when measurement noise is involved, as noisy artifacts produced by iterative methods do not appear in our case. Over various real-world 
images, we further show DDRM results on super-resolution, deblurring, inpainting and colorization (see \Cref{fig:demo}).
A DDRM trained on ImageNet also works on images that are out of its training set distribution (see \Cref{fig:ood_deblur}).
\section{Background}

\paragraph{Linear Inverse Problems.}
A general linear inverse problem is posed as
\begin{align}
    \rvy = \mH \rvx + \rvz, \label{eq:inverse-problem-def}
\end{align}
where we aim to recover the signal $\rvx \in \sR^{n}$ from measurements $\rvy \in \sR^{m}$, where $\mH \in \sR^{m \times n}$ is a known linear degradation matrix, and $\rvz \sim \gN(0, \sigma_{\rvy}^2 \mI)$ is an \textit{i.i.d.} additive Gaussian noise with known variance.
The underlying structure of $\rvx$ can be represented via a generative model, denoted as $p_\theta(\rvx)$. Given $\rvy$ and $\mH$, a posterior over the signal can be posed as: $p_\theta(\rvx | \rvy) \propto p_\theta(\rvx) p(\rvy | \rvx)$, where the ``likelihood'' term $p(\rvy | \rvx)$ is defined via \Cref{eq:inverse-problem-def}; such an approach leverages a learned prior $p_\theta(\rvx)$, and we call it an ``unsupervised'' approach based on the terminology in \cite{ongie2020deep}, as the prior does not necessarily depend on the inverse problem. Recovering $\rvx$ can be done by sampling from this posterior~\cite{bardsley2012mcmc}, which may require many iterations to produce a good sample. Alternatively, one can also approximate this posterior by learning a model via amortized inference (\textit{i.e.}, supervised learning); the model learns to predict $\rvx$ given $\rvy$, generated from $\rvx$ and a specific $\mH$. While this can be more efficient than sampling-based methods, it may generalize poorly to inverse problems that have not been trained on. 

\paragraph{Denoising Diffusion Probabilistic Models.}
Structures learned by generative models have been applied to various inverse problems and often outperform data-independent structural constraints such as sparsity~\cite{csgm}.
These generative models learn a model distribution $p_\theta(\rvx)$ that approximates a data distribution $q(\rvx)$ from samples. 
In particular, diffusion models have demonstrated impressive unconditional generative modeling performance on images~\cite{guided_diffusion}.
Diffusion models are generative models with a Markov chain structure ${\rvx_T \to \rvx_{T-1} \to \ldots \to \rvx_{1} \to \rvx_{0}}$
(where $\rvx_t \in \sR^n$), which has the following joint distribution:
$$p_\theta(\rvx_{0:T}) = p_\theta^{(T)}(\rvx_T) \prod_{t=0}^{T-1} p_\theta^{(t)}(\rvx_t | \rvx_{t+1}).$$
After drawing $\rvx_{0:T}$, only $\rvx_{0}$ is kept as the sample of the generative model.
To train a diffusion model, a fixed, factorized variational inference distribution is introduced:
$$q(\rvx_{1:T} | \rvx_0) =  q^{(T)}(\rvx_T | \rvx_0) \prod_{t=0}^{T-1}q^{(t)}(\rvx_t | \rvx_{t+1}, \rvx_0),$$
which leads to an evidence lower bound (ELBO) on the maximum likelihood objective~\cite{sohl-dickstein2015deep}.
A special property of some diffusion models is that both $p_\theta^{(t)}$ and $q^{(t)}$ are chosen as conditional Gaussian distributions for all $t < T$, and that $q(\rvx_{t} | \rvx_0)$ is also a Gaussian with known mean and covariance, \textit{i.e.}, $\rvx_{t}$ can be treated as $\rvx_0$ directly corrupted with Gaussian noise.
Thus, the ELBO objective can be reduced into the following denoising autoencoder objective (please refer to \cite{song2020denoising} for derivations):
\begin{align}
   \sum_{t=1}^{T}\gamma_t\bb{E}_{(\rvx_0, \rvx_t) \sim q(\rvx_0) q(\rvx_t | \rvx_0)}\left[\norm{\rvx_0 - f_\theta^{(t)}(\rvx_t)}_2^2 \right] \label{eq:dsm}
\end{align}
where $f^{(t)}_\theta$ is a $\theta$-parameterized neural network that aims to recover a noiseless observation from a noisy $\rvx_t$, and $\gamma_{1:T}$ are a set of positive coefficients that depend on $q(\rvx_{1:T} | \rvx_0)$. 

\section{Denoising Diffusion Restoration Models}

Inverse problem solvers based on posterior sampling often face a dilemma: unsupervised approaches apply to general problems but are inefficient, whereas supervised ones are efficient but can only address specific problems.

To solve this dilemma, we introduce Denoising Diffusion Restoration Models (DDRM),
an unsupervised solver for 
general linear inverse problems, capable of handling such tasks with or without noise in the measurements.
DDRM is efficient 
and exhibits competitive performance compared to popular unsupervised  
solvers~\cite{red, dgp, snips}.

The key idea behind DDRM is to find an unsupervised solution that also suits supervised learning objectives.
First, we describe the variational objective for DDRM over a specific inverse problem (Section~\ref{sec:3.1}). Next, we introduce specific forms of DDRM that are suitable for linear inverse problems and allow pre-trained unconditional and class-conditional diffusion models to be used directly (Sections~\ref{sec:3.2},~\ref{sec:3.3}). Finally, we discuss practical algorithms that are compute and memory efficient (Sections~\ref{sec:3.4},~\ref{sec:memory}).

\subsection{Variational Objective for DDRM}
\label{sec:3.1}
For any linear inverse problem, we define DDRM as a Markov chain ${\rvx_T \to \rvx_{T-1} \to \ldots \to \rvx_{1} \to \rvx_{0}}$ conditioned on $\rvy$, where 
$$p_\theta(\rvx_{0:T} | \rvy) = p_\theta^{(T)}(\rvx_T | \rvy) \prod_{t=0}^{T-1} p_\theta^{(t)}(\rvx_t | \rvx_{t+1}, \rvy)$$
and $\rvx_0$ is the final diffusion output.
In order to perform inference, we consider the following factorized variational distribution conditioned on $\rvy$:
$$q(\rvx_{1:T} | \rvx_0, \rvy) =  q^{(T)}(\rvx_T | \rvx_0, \rvy) \prod_{t=0}^{T-1}q^{(t)}(\rvx_t | \rvx_{t+1}, \rvx_0, \rvy),$$
leading to an ELBO objective for diffusion models conditioned on $\rvy$ (details in Appendix A).

In the remainder of the section, we construct suitable variational problems given $\mH$ and $\sigma_{\rvy}$ and connect them to unconditional diffusion generative models. To simplify notations, we will construct the variational distribution $q$ such that $q(\rvx_t | \rvx_0) = \gN(\rvx_0, \sigma_t^2 \mI)$ for noise levels $0 = \sigma_0 < \sigma_1 < \sigma_2 < \ldots < \sigma_T$.\footnote{This is called ``Variance Exploding'' in \cite{song2020sde}.} In Appendix B, we will show that this is equivalent to the distribution introduced in DDPM \cite{ddpm} and DDIM \cite{song2020denoising},\footnote{This is called ``Variance Preserving'' in \cite{song2020sde}.} up to fixed linear transformations over $\rvx_t$.

\subsection{A Diffusion Process for Image Restoration}
\label{sec:3.2}

Similar to SNIPS~\cite{snips}, we consider the singular value decomposition (SVD) of $\mH$, and perform the diffusion in its spectral space. The idea behind this is to tie the noise present in the measurements $\rvy$ with the diffusion noise in $\rvx_{1:T}$, ensuring that the diffusion result $\rvx_0$ is faithful to the measurements.
By using the SVD, we identify the data from $\rvx$ that is missing in $\rvy$, and synthesize it using a diffusion process.
In conjunction, the noisy data in $\rvy$ undergoes a denoising process. For example, in inpainting with noise (\textit{e.g.}, $\mH = \mathrm{diag}([1, \ldots, 1, 0, \ldots, 0])$, $\sigma_\rvy \geq 0$), the spectral space is simply the pixel space, so the model should generate the missing pixels and denoise the observed ones in $\rvy$.
For a general linear $\mH$, its SVD is given as 
\begin{align}
    \mH = \mU \mSigma \mV^\top 
\end{align}
where $\mU \in \sR^{m \times m}$, $\mV \in \sR^{n \times n}$ are orthogonal matrices, and $\mSigma \in \sR^{m \times n}$ is a rectangular diagonal matrix containing the singular values of $\mH$, ordered descendingly.
As this is the case in most useful degradation models, we assume $m \le n$, but our method would work for $m > n$ as well.
We denote the singular values as $s_1 \geq s_2 \geq \ldots \geq s_{m}$, and define $s_{i} = 0$ for $i \in [m+1, n]$. 

\newcommand{\bxt}[1]{\bar{\rvx}_{#1}}
\newcommand{\bxti}[2]{\bar{\rvx}_{#1}^{(#2)}}
\newcommand{\byi}[1]{\bar{\rvy}^{(#1)}}
\newcommand{\bxthetai}[1]{{\color{violet}\bar{\rvx}_{\theta, t}^{(#1)}}}

We use the shorthand notations for values in the spectral space: $\bxti{t}{i}$ is the $i$-th index of the vector $\bxt{t} = \mV^\top \rvx_t$, and $\byi{i}$ is the $i$-th index of the vector $\bar{\rvy} = \mSigma^\dagger \mU^\top \rvy$ (where $\dagger$ denotes the Moore–Penrose pseudo-inverse). 
Because $\mV$ is an orthogonal matrix, we can recover $\rvx_t$ from $\bxt{t}$ exactly by left multiplying $\mV$.
For each index $i$ in $\bxt{t}$, we define the variational distribution as:
\begin{align}
q^{(T)}(\bxti{T}{i} | \rvx_0, \rvy) = & \begin{cases}
\gN(\byi{i}, \sigma_{T}^2 - \frac{\sigma_{\rvy}^2}{s_i^2}) & \text{if } s_i > 0 \\
\gN(\bxti{0}{i}, \sigma_{T}^2) & \text{if } s_i = 0
\end{cases} \label{eq:qt-init} \\
    q^{(t)}(\bxti{t}{i} | \rvx_{t+1}, \rvx_0, \rvy) = 
    & \begin{cases}
     \gN(\bxti{0}{i} + \sqrt{1 - \eta^2} \sigma_{t} \frac{\bxti{t+1}{i} - \bxti{0}{i}}{\sigma_{t+1}},\eta^2 \sigma_{t}^2) & \text{if } s_i = 0
    \\
    \gN(\bxti{0}{i} + \sqrt{1 - \eta^2} \sigma_{t} \frac{\byi{i} - \bxti{0}{i}}{\sigma_{\rvy} / s_i}, \eta^2 \sigma_{t}^2) & \text{if } \sigma_{t} < \frac{\sigma_{\rvy}}{s_i} \\
    \gN((1 - \eta_b) \bxti{0}{i} + \eta_b \byi{i}, \sigma_{t}^2 - \frac{\sigma_{\rvy}^2}{s_i^2} \eta_b^2) & \text{if } \sigma_{t} \geq \frac{\sigma_{\rvy}}{s_i}
    \end{cases} \label{eq:qt-transition}
\end{align}
where $\eta \in (0, 1]$ is a hyperparameter controlling the variance of the transitions, and $\eta$ and $\eta_b$ may depend on
$\sigma_t, s_i, \sigma_{\rvy}$. We further assume that $\sigma_T \geq \sigma_{\rvy} / s_i$ for all positive $s_i$.\footnote{This assumption is fair, as we can set a sufficiently large $\sigma_T$.}

In the following statement, we show that this construction has the ``Gaussian marginals'' property similar to the inference distribution used in unconditional diffusion models~\cite{ddpm}. 

\begin{restatable}{proposition}{consistency}
\label{thm:consistency}
The conditional distributions $q^{(t)}$ defined in Equations~\ref{eq:qt-init} and~\ref{eq:qt-transition} satisfy the following:
\begin{align}
    q(\rvx_t | \rvx_0) = \gN(\rvx_0, \sigma_t^2 \mI),
\end{align}
defined by marginalizing over $\rvx_{t'}$ (for all $t' > t$) and $\rvy$, where $q(\rvy | \rvx_0)$ is defined as in \Cref{eq:inverse-problem-def} with $\rvx = \rvx_0$. 
\end{restatable}
We place the proof in Appendix C.
\newcommand*{\propositionCrefname}{Proposition}
Intuitively, our construction considers different cases for each index of the spectral space. (\emph{i}) If the corresponding singular value is zero, then $\rvy$ does not directly provide any information to that index, and the update is similar to regular unconditional generation. (\emph{ii}) If the singular value is non-zero, then the updates consider the information provided by $\rvy$, which further depends on whether the measurements' noise level in the spectral space ($\sigma_{\rvy} / s_i$) is larger than the noise level in the diffusion model ($\sigma_t$) or not; the measurements in the spectral space $\byi{i}$ are then scaled differently for these two cases in order to ensure \Cref{thm:consistency} holds.

Now that we have defined $q^{(t)}$ as a series of Gaussian conditionals, we define our model distribution $p_\theta$ as a series of Gaussian conditionals as well. Similar to DDPM, we aim to obtain predictions of $\rvx_0$ at every step $t$; and to simplify notations, we use the symbol {\color{violet} $\rvx_{\theta, t}$} to represent this prediction made by a model\footnote{Equivalently, the authors of \cite{ddpm} predict the noise values to subtract in order to recover {\color{violet}$\rvx_{\theta, t}$}.} $f_\theta(\rvx_{t+1}, t+1): \sR^{n} \times \sR \to \sR^{n}$ that takes in the sample $\rvx_{t+1}$ and the conditioned time step $(t+1)$. 
We also define ${\color{violet}\bar{\rvx}_{\theta, t}^{(i)}}$ as the $i$-th index of ${\color{violet}{\bar{\rvx}_{\theta, t}}} = \mV^\top {\color{violet}\rvx_{\theta, t}}$. 

We define DDRM with trainable parameters $\theta$ as follows:
\begin{align}
p_\theta^{(T)}(\bxti{T}{i} | \rvy) = & \begin{cases}
\gN(\byi{i}, \sigma_{T}^2 - \frac{\sigma_{\rvy}^2}{s_i^2}) & \text{if } s_i > 0 \\
\gN(0, \sigma_{T}^2) & \text{if } s_i = 0
\end{cases} \label{eq:pt-init} \\
    p_\theta^{(t)}(\bxti{t}{i} | \rvx_{t+1}, \rvy) =
    & \begin{cases}
     \gN(\bxthetai{i} + \sqrt{1 - \eta^2} \sigma_{t} \frac{\bxti{t+1}{i} - \bxthetai{i}}{\sigma_{t+1}},\eta^2 \sigma_{t}^2) & \text{if } s_i = 0
    \\
    \gN(\bxthetai{i} + \sqrt{1 - \eta^2} \sigma_{t} \frac{\byi{i} - \bxthetai{i}}{\sigma_{\rvy} / s_i}, \eta^2 \sigma_{t}^2) & \text{if } \sigma_{t} < \frac{\sigma_{\rvy}}{s_i} \\
    \gN((1 - \eta_b) \bxthetai{i} + \eta_b \byi{i}, \sigma_{t}^2 - \frac{\sigma_{\rvy}^2}{s_i^2} \eta_b^2) & \text{if } \sigma_{t} \geq \frac{\sigma_{\rvy}}{s_i}.
    \end{cases} \label{eq:pt-transition}
\end{align}

Compared to $q^{(t)}$ in \Cref{eq:qt-init,eq:qt-transition}, our definition of $p_\theta^{(t)}$ merely replaces $\bxti{0}{i}$ (which we do not know at sampling) with $\bxthetai{i}$ (which depends on our predicted ${\color{violet}\rvx_{\theta, t}}$) when $t < T$, and replaces $\bxti{0}{i}$ with $0$ when $t = T$. It is possible to learn the variances~\cite{nichol2021improved} or consider alternative constructions where \Cref{thm:consistency} holds; we leave these options as future work.

\subsection{``Learning'' Image Restoration Models}\label{sec:3.3}
Once we have defined $p_\theta^{(t)}$ and $q^{(t)}$ by choosing $\sigma_{1:T}$, $\eta$ and $\eta_b$, we can learn model parameters $\theta$ by maximizing the resulting ELBO objective (in Appendix A). However, this approach is not desirable since we have to learn a different model for each inverse problem (given $\mH$ and $\sigma_{\rvy}$), which is not flexible enough for arbitrary inverse problems. Fortunately, this does not have to be the case.
In the following statement, we show that an optimal solution to DDPM / DDIM can also be an optimal solution to a DDRM problem, under reasonable assumptions used in prior work~\cite{ddpm,song2020denoising}.
\begin{restatable}{theorem}{ddrmmain}
\label{thm:ddrm-main}
Assume that the models $f_\theta^{(t)}$ and $f_\theta^{(t')}$ do not have weight sharing whenever $t \neq t'$,
then when $\eta = 1$ and 
$\eta_b = \frac{2\sigma_t^2}{\sigma_t^2  + \sigma_{\rvy}^2 / s_i^2}$,
the ELBO objective of DDRM (details in Appendix A) can be rewritten in the form of the DDPM / DDIM objective in \Cref{eq:dsm}.
\end{restatable}
We place the proof in Appendix C.

Even for different choices of $\eta$ and $\eta_b$, the proof shows that the DDRM objective is a weighted sum-of-squares error in the spectral space, and thus pre-trained DDPM models are good approximations to the optimal solution.
Therefore, we can apply the same diffusion model (\textit{unconditioned} on the inverse problem) using the updates in \Cref{eq:pt-init} and \Cref{eq:pt-transition} and only modify $\mH$ and its SVD ($\mU$, $\mSigma$, $\mV$) for various linear inverse problems.

\subsection{Accelerated Algorithms for DDRM}\label{sec:3.4}
Typical diffusion models are trained with many timesteps (\textit{e.g.}, 1000) to achieve optimal unconditional image synthesis quality, but sampling speed is slow as many NFEs are required. Previous works~\cite{song2020denoising,guided_diffusion} have accelerated this process by ``skipping'' steps with appropriate update rules. This is also true for DDRM, since we can obtain the denoising autoencoder objective in \Cref{eq:dsm} for any choice of increasing $\sigma_{1:T}$. For a pre-trained diffusion model with $T'$ timesteps, we can choose $\sigma_{1:T}$ to be a subset of the $T'$ steps used in training.

\subsection{Memory Efficient SVD}\label{sec:memory}
Our method, similar to SNIPS~\cite{snips}, utilizes the SVD of the degradation operator $\mH$. This constitutes a memory consumption bottleneck in both algorithms as well as other methods such as Plug and Play (PnP) ~\cite{pnp}, as storing the matrix $\mV$ has a space complexity of $\Theta(n^2)$ for signals of size $n$.
By leveraging special properties of the matrices $\mH$ used, we can reduce this complexity to $\Theta(n)$ for denoising, inpainting, super resolution, deblurring, and colorization (details in Appendix D).

\section{Related Work}
\begin{figure}
\begin{subfigure}{0.49\textwidth}
    \centering
    \def\arraystretch{0.7}
    \setlength\tabcolsep{0.05cm}
    \begin{tabular}{l cccc}
        \rotatebox{90}{\hspace{0.35cm}\tiny Original}
        & \includegraphics[width=1.5cm,height=1.5cm]{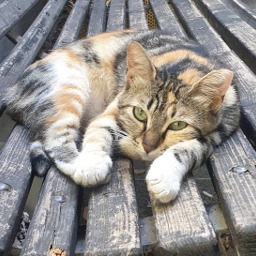}
        & \includegraphics[width=1.5cm,height=1.5cm]{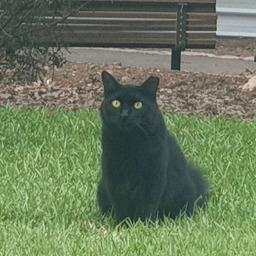}
        & \includegraphics[width=1.5cm,height=1.5cm]{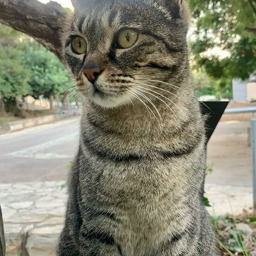}
        & \includegraphics[width=1.5cm,height=1.5cm]{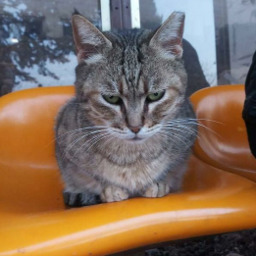}
        \\
        \rotatebox{90}{\hspace{0.3cm} \tiny Degraded}
        & \includegraphics[width=1.5cm,height=1.5cm]{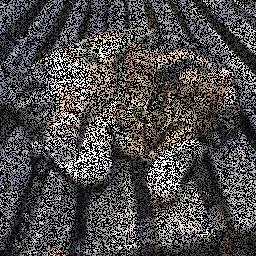}
        & \includegraphics[width=1.5cm,height=1.5cm]{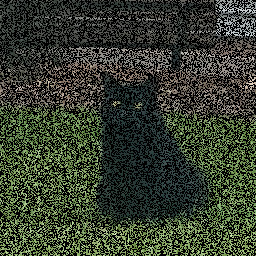}
        & \includegraphics[width=1.5cm,height=1.5cm]{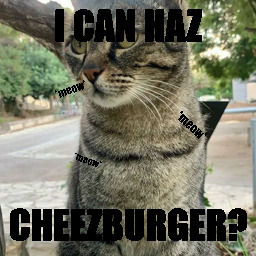}
        & \includegraphics[width=1.5cm,height=1.5cm]{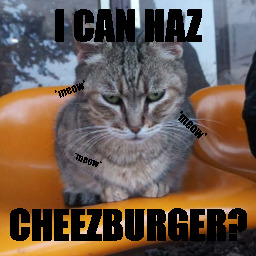}
        \\
        \rotatebox{90}{\hspace{0.12cm} \tiny DDRM ($20$)}
        & \includegraphics[width=1.5cm,height=1.5cm]{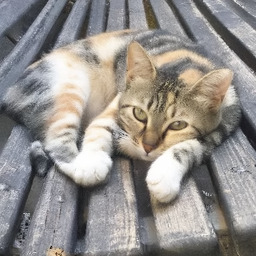}
        & \includegraphics[width=1.5cm,height=1.5cm]{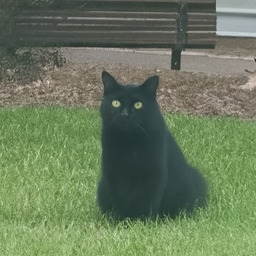}
        & \includegraphics[width=1.5cm,height=1.5cm]{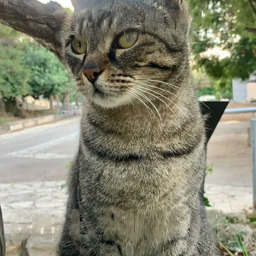}& \includegraphics[width=1.5cm,height=1.5cm]{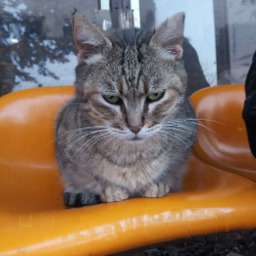}
    \end{tabular}
    \caption{Inpainting results on cat images.}
\end{subfigure}
~
\begin{subfigure}{0.49\textwidth}
    \centering
    \def\arraystretch{0.7}
    \setlength\tabcolsep{0.05cm}
    \begin{tabular}{l cccc}
        \rotatebox{90}{\hspace{0.35cm}\tiny Original}
        & \includegraphics[width=1.5cm,height=1.5cm]{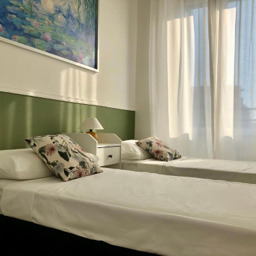}
        & \includegraphics[width=1.5cm,height=1.5cm]{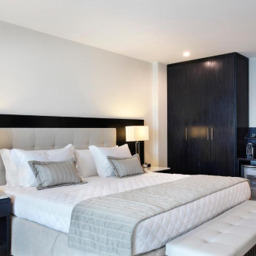}
        & \includegraphics[width=1.5cm,height=1.5cm]{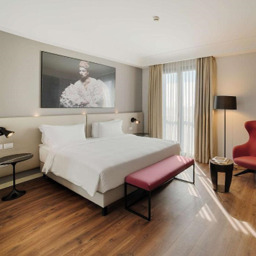}
        & \includegraphics[width=1.5cm,height=1.5cm]{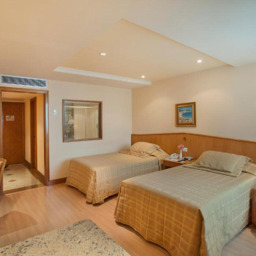}
        \\
        \rotatebox{90}{\hspace{0.3cm}\tiny Degraded}
        & \includegraphics[width=1.5cm,height=1.5cm]{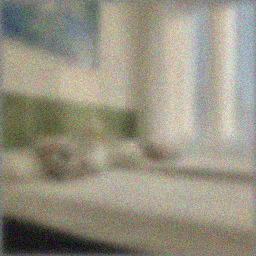}
        & \includegraphics[width=1.5cm,height=1.5cm]{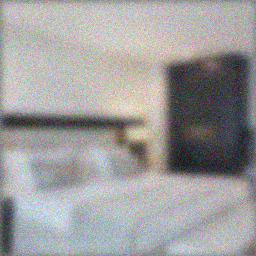}
        & \includegraphics[width=1.5cm,height=1.5cm]{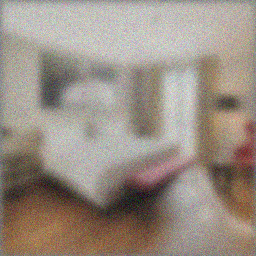}
        & \includegraphics[width=1.5cm,height=1.5cm]{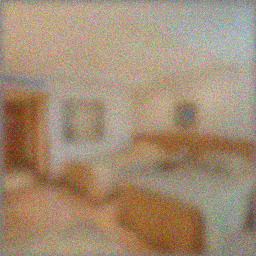}
        \\
        \rotatebox{90}{\hspace{0.12cm}\tiny DDRM ($20$)}
        & \includegraphics[width=1.5cm,height=1.5cm]{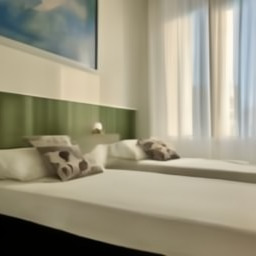}
        & \includegraphics[width=1.5cm,height=1.5cm]{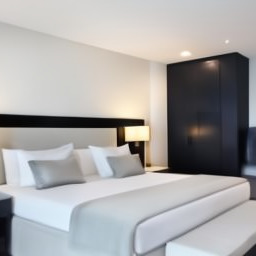}
        & \includegraphics[width=1.5cm,height=1.5cm]{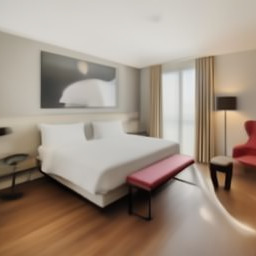}
        & \includegraphics[width=1.5cm,height=1.5cm]{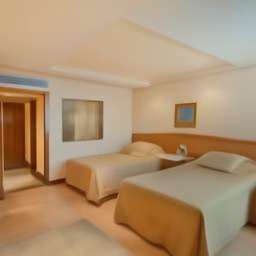}
    \end{tabular}
    \caption{Deblurring results ($\sigma_{\rvy} = 0.05$) on bedroom images.
    }
\end{subfigure}
\caption{DDRM results on bedroom and cat images, for inpainting and deblurring.}
\label{fig:bed_deblur_cat_inp}
\end{figure}

Various deep learning solutions have been suggested for solving inverse problems under different settings (see a detailed survey in \cite{survey}). We focus on the \emph{unsupervised} setting, where we have access to a dataset of clean images at training time, but the degradation model is known only at inference time. This setup is inherently general to all linear inverse problems, a property desired in many real-world applications such as medical imaging \cite{song2021medical,jalal2021robust}.

Almost all unsupervised inverse problem solvers utilize a trained neural network in an iterative scheme.
PnP, RED, and their successors \cite{pnp, red, deepred, sun2019online} apply a denoiser as part of an iterative optimization algorithm such as steepest descent, fixed-point, or alternating direction method of multipliers (ADMM). OneNet \cite{onenet} trained a network to directly learn the proximal operator of ADMM.  
A similar use of denoisers in different iterative algorithms is proposed in \cite{ldamp, agem, laumont}. The authors of \cite{santurkar2019image} leverages robust classifiers learned with additional class labels.

Another approach is to search the latent space of a generative model for a generated image that, when degraded, is as close as possible to the given measurements. Multiple such methods were suggested, mainly focusing on generative adversarial networks (GANs) \cite{csgm, ilo, pulse}. While they exhibit impressive results on images of a specific class, most notably face images, these methods are not shown to be largely successful under a more diverse dataset such as ImageNet \cite{imagenet}. Deep Generative Prior (DGP) mitigates this issue by optimizing the latent input as well as the weights of the GAN's generator \cite{dgp}.

More recently, denoising diffusion models were used to solve inverse problems in both supervised (\textit{i.e.}, degradation model is known during training) \cite{saharia2021image,palette,guided_diffusion, chung2021come, whang2021deblurring} and unsupervised settings \cite{simoncelli, sid, snips, instance_optimal, song2021medical, song2020sde, ilvr}. Unlike previous approaches, most diffusion-based methods can successfully recover images from measurements with significant noise. However, these methods are very slow, often requiring hundreds or thousands of iterations, and are yet to be proven on diverse datasets. Our method, motivated by variational inference, obtains problem-specific, non-equilibrium update rules that lead to high-quality solutions in much fewer iterations.

ILVR~\cite{ilvr} suggests a diffusion-based method that handles noiseless super-resolution, and can run in $250$ steps. In Appendix H, we prove that when applied on the same underlying generative diffusion model, ILVR is a special case of DDRM.
Therefore, ILVR can be further accelerated to run in $20$ steps, but unlike DDRM, it provides no clear way of handling noise in the measurements.
Similarly, the authors of~\cite{simoncelli} suggest a score-based solver for inverse problems that can converge in a small number of iterations, but does not handle noise in the measurements.
\section{Experiments}

\begin{table}
    \centering
    \caption{Noiseless $4\times$ super-resolution and deblurring results on ImageNet 1K ($256\times256$).}
    \label{tab:in1k_noiseless}
    \begin{center}
    \begin{tabular}{l p{0.0001\textwidth} cccc p{0.0001\textwidth} cccc}
        \toprule
        \multirow{2}{*}{Method} & ~ &  \multicolumn{4}{c}{$4\times$ super-resolution} & ~ & \multicolumn{4}{c}{Deblurring} \\
        & ~ & PSNR$\uparrow$ & SSIM$\uparrow$ & KID$\downarrow$ & NFEs$\downarrow$ & ~ & PSNR$\uparrow$ & SSIM$\uparrow$ & KID$\downarrow$ & NFEs$\downarrow$ \\
        \midrule
        Baseline & ~ & $25.65$ & $0.71$ & $44.90$ & \first{0} & ~ & $19.26$ & $0.48$ & $38.00$ & \first{0} \\
        DGP & ~ & $23.06$ & $0.56$ & $21.22$ & $1500$ & ~ & $22.70$ & $0.52$ & $27.60$ & $1500$ \\
        RED & ~ & \second{26.08} & $0.73$ & $53.55$ & $100$ & ~ & $26.16$ & $0.76$ & $21.21$ & $500$ \\
        SNIPS & ~ & $17.58$ & $0.22$ & $35.17$ & $1000$ & ~ & $34.32$ & $0.87$ & \first{0.49} & $1000$ \\
        \midrule
        DDRM & ~ & \first{26.55} & \second{0.72} & \second{7.22} & \second{20} & ~ & \second{35.64} & \second{0.95} & $0.71$ & \second{20} \\ 
        DDRM-CC & ~ & \first{26.55} & \first{0.74} & \first{6.56} & \second{20} & ~ & \first{35.65} & \first{0.96} & \second{0.70} & \second{20} \\ 
        \bottomrule
    \end{tabular}
    \end{center}
\end{table}

\subsection{Experimental Setup}
We demonstrate our algorithm's capabilities using the diffusion models from \cite{ddpm}, which are trained on CelebA-HQ \cite{celeba_hq}, LSUN bedrooms, and LSUN cats \cite{lsun}  (all $256 \times 256$ pixels).
We test these models on images from FFHQ \cite{stylegan}, and pictures from the internet of the considered LSUN category, respectively.
In addition, we use the models from \cite{guided_diffusion}, trained on the training set of ImageNet $256 \times 256$ and $512 \times 512$, and tested on the corresponding validation set.
Some of the ImageNet models require class information. For these models, we use the ground truth labels as input, and denote our algorithm as DDRM class conditional (DDRM-CC).
In all experiments, we use $\eta = 0.85$, $\eta_b = 1$, and a uniformly-spaced timestep schedule based on the 1000-step pre-trained models (more details in Appendix E). The number of NFEs (timesteps) is reported in each experiment.

In each of the inverse problems we show,
pixel values are in the range $[0, 1]$,
and the degraded measurements are obtained as follows:
(\emph{i}) for super-resolution, we use a block averaging filter to downscale the images by a factor of $2$, $4$, or $8$ in each axis; 
(\emph{ii}) for deblurring, the images are blurred by a $9 \times 9$ uniform kernel, and singular values below a certain threshold are zeroed, making the problem more ill-posed.
(\emph{iii}) for colorization, the grayscale image is an average of the red, green, and blue channels of the original image;
(\emph{iv}) and for inpainting, we mask parts of the original image with text overlay or randomly drop $50\%$ of the pixels. Additive white Gaussian noise can optionally be added to the measurements in all inverse problems.
We additionally conduct experiments on bicubic super-resolution and deblurring with an anisotropic Gaussian kernel in Appendix I.

Our code is available at \url{https://github.com/bahjat-kawar/ddrm}.

\begin{table}
    \centering
    \caption{$4\times$ super resolution and deblurring results on ImageNet 1K ($256\times256$). Input images have an additive noise of $\sigma_{\rvy} = 0.05$.}
    \label{tab:in1k_noisy}
    \begin{center}
    \begin{tabular}{l p{0.0001\textwidth} cccc p{0.0001\textwidth} cccc}
        \toprule
        \multirow{2}{*}{Method} & ~ &  \multicolumn{4}{c}{$4\times$ super-resolution} & ~ & \multicolumn{4}{c}{Deblurring} \\
        & ~ & PSNR$\uparrow$ & SSIM$\uparrow$ & KID$\downarrow$ & NFEs$\downarrow$ & ~ & PSNR$\uparrow$ & SSIM$\uparrow$ & KID$\downarrow$ & NFEs$\downarrow$ \\
        \midrule
        Baseline & ~ & $22.55$ & $0.46$ & $67.86$ & \first{0} & ~ & $18.35$ & $0.20$ & $75.50$ & \first{0} \\
        DGP & ~ & $20.69$ & $0.43$ & $42.17$ & $1500$ & ~ & $21.20$ & $0.45$ & $34.02$ & $1500$ \\
        RED & ~ & $22.90$ & $0.49$ & $43.45$ & $100$ & ~ & $14.69$ & $0.08$ & $121.82$ & $500$ \\
        SNIPS & ~ & $16.30$ & $0.14$ & $67.77$ & $1000$ & ~ & $16.37$ & $0.14$ & $77.96$ & $1000$ \\
        \midrule
        DDRM & ~ & \second{25.21} & \second{0.66} & \second{12.43} & \second{20} & ~ & \second{25.45} & \second{0.66} & \second{15.24} & \second{20} \\ 
        DDRM-CC & ~ & \first{25.22} & \first{0.67} & \first{10.82} & \second{20} & ~ & \first{25.46} & \first{0.67} & \first{13.49} & \second{20} \\ 
        \bottomrule
    \end{tabular}
    \end{center}
\end{table}

\newcommand{\imidx}[0]{4}
\begin{figure}
    \centering
    \def\arraystretch{0.7}
    \setlength\tabcolsep{0.05cm}
    \begin{tabular}{ccc ccc}
        \includegraphics[width=2.2cm,height=2.2cm]{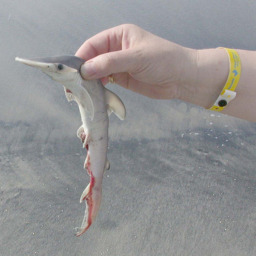}
        & \includegraphics[width=2.2cm,height=2.2cm]{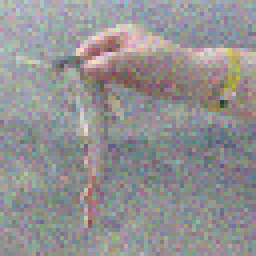}
        & \includegraphics[width=2.2cm,height=2.2cm]{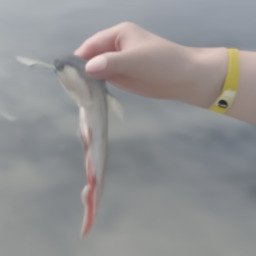}
        & \includegraphics[width=2.2cm,height=2.2cm]{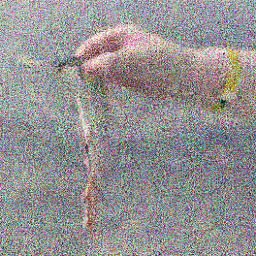}
        & \includegraphics[width=2.2cm,height=2.2cm]{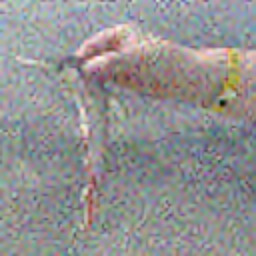}
        & \includegraphics[width=2.2cm,height=2.2cm]{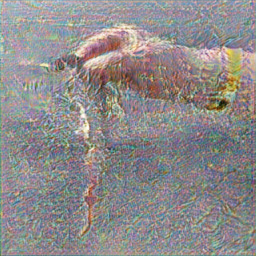}
        \\
        {Original} & {Low-res} & {DDRM ($20$)}
        & {SNIPS} & {RED} & {DGP}
    \end{tabular}
    \caption{$4\times$ noisy super resolution comparison with $\sigma_{\rvy}=0.05$.}
    \label{fig:sr4_comp}
\end{figure}

\renewcommand{\imwidth}{2.2cm}
\begin{figure*}
    \centering
    \def\arraystretch{0.7}
    \setlength\tabcolsep{0.03cm}
    \begin{tabular}{lcccccc}
        \rotatebox{90}{\hspace{0.55cm}`teapot'} &
        \includegraphics[width=\imwidth,height=\imwidth]{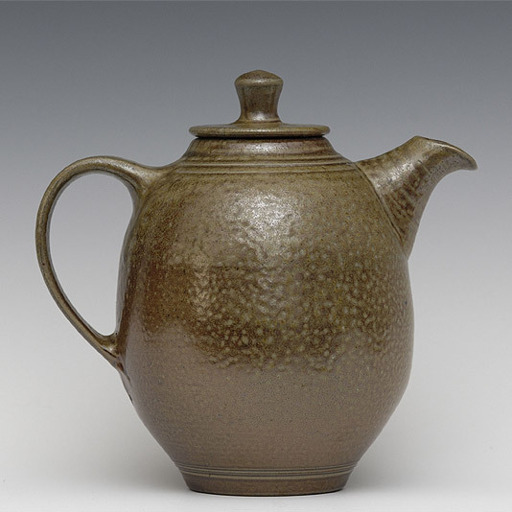}
        & \includegraphics[width=\imwidth,height=\imwidth]{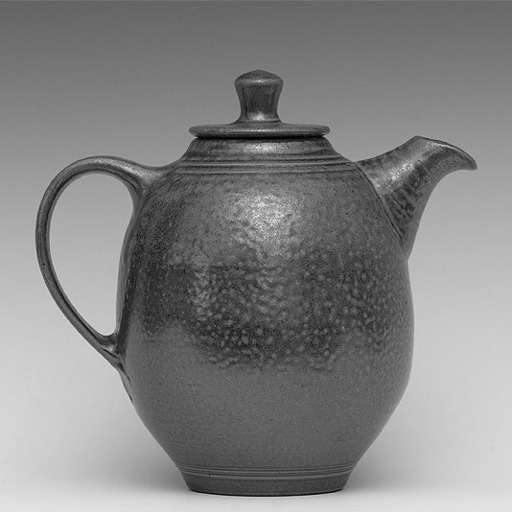}& \hspace{0.2cm}
        \forloop{row}{1}{\value{row} < 4}{
            & \includegraphics[width=\imwidth,height=\imwidth]{./figures/imagenet_color/teapot/0_s\arabic{row}_-1.jpg}
        } \\
        \rotatebox{90}{\hspace{0.3cm}`brown bear'} &
        \includegraphics[width=\imwidth,height=\imwidth]{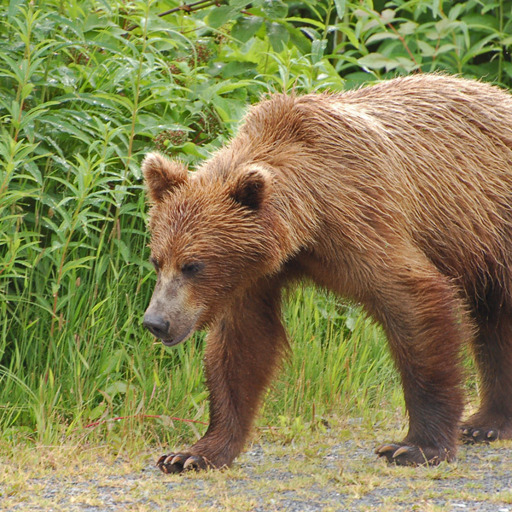}
        & \includegraphics[width=\imwidth,height=\imwidth]{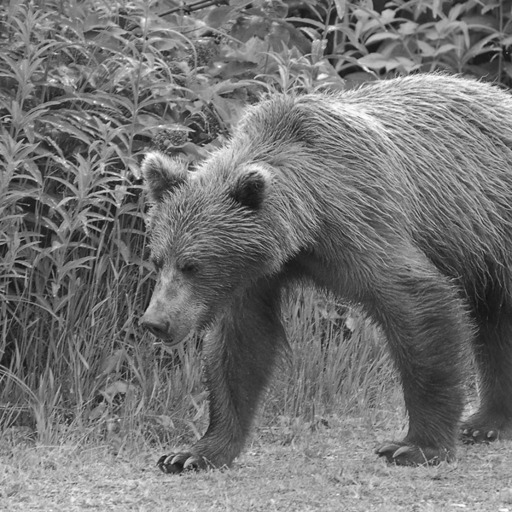} & \hspace{0.2cm}
        \forloop{row}{1}{\value{row} < 4}{
            & \includegraphics[width=\imwidth,height=\imwidth]{./figures/imagenet_color/brown_bear/144_s\arabic{row}_-1.jpg}
        } \\
        & Original & Grayscale & \hspace{0.2cm} &
        \multicolumn{3}{c}{Samples from DDRM-CC ($100$)}
    \end{tabular}
    \caption{$512 \times 512$ ImageNet colorization. DDRM-CC produces various samples for multiple runs on the same input.
    }
    \label{fig:imagenet_color}
\end{figure*}

\subsection{Quantitative Experiments}
In order to quantify DDRM's performance, we focus on the ImageNet dataset ($256 \times 256$) for its diversity. For each experiment, we report the average peak signal-to-noise ratio (PSNR)
 and structural similarity index measure (SSIM)~\cite{ssim} 
to measure faithfulness to the original image, and the kernel Inception distance (KID) \cite{kid}, multiplied by $10^3$, to measure the resulting image quality.

We compare DDRM (with $20$ and $100$ steps) with other unsupervised methods that work in reasonable time (requiring $1500$ NFEs or less) and can operate on ImageNet. Namely, we compare with RED \cite{red}, DGP \cite{dgp}, and SNIPS \cite{snips}. The exact setup of each method is detailed in Appendix F.
We used the same hyperparameters for noisy and noiseless versions of the same problem for DGP, RED, and SNIPS, as tuning them for each version would compromise their unsupervised nature.
Nevertheless, the performance of baselines like RED with such a tuning does not surpass that of DDRM, as we show in Appendix F.
In addition, we show upscaling by bicubic interpolation as a baseline for super-resolution, and the blurry image itself as a baseline for deblurring.
OneNet \cite{onenet} is not included in the comparisons as it is limited to images of size $64 \times 64$, and generalization to higher dimensions requires an improved network architecture.

We evaluate all methods on the problems of $4\times$ super-resolution and deblurring, on one validation set image from each of the $1000$ ImageNet classes, following \cite{dgp}.
Table \ref{tab:in1k_noiseless} shows that DDRM outperforms all baseline methods, in all metrics, and on both problems with only $20$ steps.
The only exception
to this is that SNIPS achieves better KID than DDRM in noiseless deblurring, but it requires $50\times$ more NFEs to do so.
Note that the runtime of all the tested methods is perfectly linear with NFEs, with negligible differences in time per iteration.
DGP and DDRM-CC use ground-truth class labels for the test images to aid in the restoration process, and thus have an unfair advantage.

DDRM's appeal compared to previous methods becomes more substantial when significant noise is added to the measurements.
Under this setting, DGP, RED, and SNIPS all fail to produce viable results, as evident in Table \ref{tab:in1k_noisy} and \Cref{fig:sr4_comp}.
Since DDRM is fast, we also evaluate it on the entire ImageNet validation set in Appendix F.

\subsection{Qualitative Experiments}
\begin{figure}
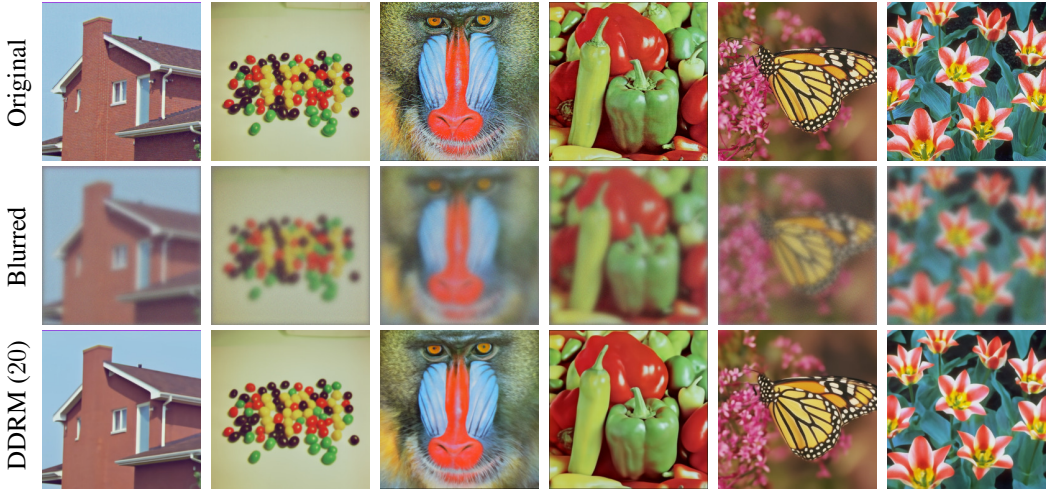

    \centering
    \def\arraystretch{0.7}
    \setlength\tabcolsep{0.03cm}
    \begin{tabular}{lcccccc}
        \rotatebox{90}{\hspace{0.45cm}Original}
        \forloop{row}{0}{\value{row} < 6}{
            & \includegraphics[width=2.1cm,height=2.1cm]{./figures/ood_deblur/orig_\arabic{row}.jpg}
        } \\
        \rotatebox{90}{\hspace{0.5cm}Blurred}
        \forloop{row}{0}{\value{row} < 6}{
            & \includegraphics[width=2.1cm,height=2.1cm]{./figures/ood_deblur/y0_\arabic{row}.jpg}
        } \\
        \rotatebox{90}{\hspace{0.25cm}DDRM ($20$)}
        \forloop{row}{0}{\value{row} < 6}{
            & \includegraphics[width=2.1cm,height=2.1cm]{./figures/ood_deblur/\arabic{row}_s0_-1.jpg}
        }
    \end{tabular}
    \caption{Results on $256 \times 256$ USC-SIPI images using an ImageNet model.
    Blurred images have a noise of $\sigma_\rvy = 0.01$.}
    \label{fig:ood_deblur}
\end{figure}

DDRM produces high quality reconstructions across all the tested datasets and problems, as can be seen in Figures \ref{fig:demo} and \ref{fig:bed_deblur_cat_inp}, and in Appendix I.
As it is a posterior sampling algorithm, DDRM can produce multiple outputs for the same input, as demonstrated in \Cref{fig:imagenet_color}.
Moreover, the unconditional ImageNet diffusion models can be used to solve inverse problems on out-of-distribution images with general content. In \Cref{fig:ood_deblur}, we show DDRM successfully restoring $256 \times 256$ images from USC-SIPI \cite{sipi} that do not necessarily belong to any ImageNet class (more results in Appendix I).

\section{Conclusions}

We have introduced DDRM, a general sampling-based linear inverse problem solver based on unconditional/class-conditional diffusion generative models as learned priors. Motivated by variational inference, DDRM only requires a few number of NFEs (\textit{e.g.}, 20) compared to other sampling-based baselines (\textit{e.g.}, 1000 for SNIPS) and achieves scalability in multiple useful scenarios, including denoising, super-resolution, deblurring, inpainting, and colorization. We demonstrate the empirical successes of DDRM on various problems and datasets, including general natural images outside the distribution of the observed training set.
To our best knowledge, DDRM is the first unsupervised method that effectively and efficiently samples from the posterior distribution of inverse problems with significant noise, and can work on natural images with general content.

In terms of future work, apart from further optimizing the timestep and variance schedules, 
it would be interesting to investigate the following: (\textit{i}) applying DDRM to non-linear inverse problems, (\textit{ii}) addressing scenarios where the degradation operator is unknown, and (\textit{iii}) self-supervised training techniques inspired by DDRM as well as ones used in supervised techniques~\cite{palette} that further improve performance of unsupervised models for image restoration.

\section*{Acknowledgements}
We thank Kristy Choi, Charlie Marx, and Avital Shafran for insightful discussions and feedback. This research was supported by NSF (\#1651565, \#1522054, \#1733686), ONR (N00014-19-1-2145), AFOSR (FA9550-19-1-0024), ARO (W911NF-21-1-0125), Sloan Fellowship, Amazon AWS, Stanford Institute for Human-Centered Artificial Intelligence (HAI), Google Cloud, the Israel Science Foundation (ISF) under Grant 335/18, the Israeli Council For Higher Education - Planning \& Budgeting Committee, and the Stephen A. Kreynes Fellowship. 

\bibliography{refs}



\newpage
\appendix
\section{Details of the DDRM ELBO objective}
\label{app:sec:details}

DDRM is a Markov chain conditioned on $\rvy$, which would lead to the following ELBO objective~\cite{song2020denoising}:
\begin{align}
    & \bb{E}_{\rvx_0 \sim q(\rvx_0), \rvy \sim q(\rvy | \rvx_0)}[\log p_\theta(\rvx_0 | \rvy)]  \\
    \geq \ & - \bb{E}\left[\sum_{t=1}^{T-1} \KL(q^{(t)}(\rvx_t | \rvx_{t+1}, \rvx_{0}, \rvy) \Vert p_\theta^{(t)}(\rvx_t | \rvx_{t+1}, \rvy))\right] + \bb{E}\left[\log p_\theta^{(0)}(\rvx_0 | \rvx_{1}, \rvy)\right] \nonumber \\
    & - \bb{E}[\KL(q^{(T)}(\rvx_T | \rvx_0, \rvy) \Vert p_\theta^{(T)}(\rvx_T | \rvy))]\label{eq:elbo}
\end{align}
where $q(\rvx_0)$ is the data distribution, $q(\rvy | \rvx_0)$ follows Equation 1 in the main paper, the expectation on the right hand side is given by sampling $\rvx_0 \sim q(\rvx_0)$, $\rvy \sim q(\rvy | \rvx_0)$, $\rvx_{T} \sim q^{(T)}(\rvx_T | \rvx_0, \rvy)$, and $\rvx_{t} \sim q^{(t)}(\rvx_t | \rvx_{t+1}, \rvx_0, \rvy)$ for $t \in [1, T-1]$.

\section{Equivalence between ``Variance Preserving'' and ``Variance Exploding'' Diffusion Models}
\label{app:sec:ve-vp}
\newcommand{\grvx}[1]{{\color{teal}\rvx_{#1}}}

In our main paper, we describe our methods based on the ``Variance Exploding'' hyperparameters $\sigma_t$, where $\sigma_t \in [0, \infty)$ and
\begin{align}
    q(\rvx_t | \rvx_0) = \gN(\rvx_0, \sigma_t^2 \mI).
\end{align}
In DDIM~\cite{song2020denoising}, the hyperparameters are ``Variance Preserving'' ones $\alpha_t$, where $\alpha_t \in (0, 1]$ and
\begin{align}
    q(\grvx{t} | \rvx_0) = \gN(\sqrt{\alpha_t} \rvx_0, (1 - \alpha_t) \mI).
\end{align}
We use the colored notation $\grvx{t}$ to emphasize that this is different from $\rvx_t$ (an exception is $\grvx{0} = \rvx_0$). Using the reparametrization trick, we have that:
\begin{gather}
    \rvx_t = \rvx_0 + \sigma_t \epsilon  \label{eq:ve-reparam} \\
    \grvx{t} = \sqrt{\alpha_t} \rvx_0 + \sqrt{1 - \alpha_t} \epsilon
\end{gather}
where $\epsilon \sim \gN(0, \mI)$. We can divide by $\sqrt{1 + \sigma_t^2}$ in both sides of \autoref{eq:ve-reparam}:
\begin{gather}
    \frac{\rvx_t}{\sqrt{1 + \sigma_t^2}} = \frac{\rvx_0}{\sqrt{1 + \sigma_t^2}} + \frac{\sigma_t}{\sqrt{1 + \sigma_t^2}} \epsilon. \label{eq:ve-reparam2}
\end{gather}
Let $\alpha_t = 1 / (1 + \sigma_t^2)$, and let $\grvx{t} = \rvx_t / \sqrt{1 + \sigma_t^2}$; then from \autoref{eq:ve-reparam2} we have that
\begin{gather}
    \grvx{t} = \sqrt{\alpha_t} \rvx_0 + \sqrt{1 - \alpha_t} \epsilon,
\end{gather}
which is equivalent to the ``Variance Preserving'' case. Therefore, we can use ``Variance Preserving'' models, such as DDPM, directly in our DDRM updates, even though the latter uses the ``Variance Exploding'' parametrization:
\begin{enumerate}
    \item From $\grvx{t}$, obtain predictions $\epsilon$ and $\rvx_t = \grvx{t} \sqrt{1 + \sigma_t^2}$.
    \item From $\rvx_{t}$ and $\epsilon$, apply DDRM updates to get $\rvx_{t-1}$.
    \item From $\rvx_{t-1}$, get $\grvx{t-1} = \rvx_{t-1} / \sqrt{1 + \sigma_{t-1}^2}$.
\end{enumerate}

Note that although the inference algorithms are shown to be equivalent, the choice between "Variance Preserving" and "Variance Exploding" may affect the training of diffusion networks.

\section{Proofs}
\label{app:sec:proofs}
\consistency*

\begin{proof}
The proof uses a basic property of Gaussian marginals (see \cite{bishop2006pattern} for the complete version). 
\begin{enumerate}
\item If $p(z_1 | z_0) = \gN(z_0, V_1)$, $p(z_2 | z_1) = \gN(\alpha z_1, V_2)$, then $p(z_2 | z_0) = \gN(\alpha z_0, \alpha^2 V_1 + V_2)$.
 \item If $p(z_1) = \gN(\mu_1, V_1)$ and $p(z_2) = \gN(\mu_2, V_2)$, then $p(z_1 + z_2) = \gN(\mu_1 + \mu_2, V_1 + V_2)$.
\end{enumerate}

First, we note that $q(\rvy | \rvx_0)$ is defined from Equation 1 in the main paper, and thus for all $i$:
\begin{align}
    q(\byi{i} | \rvx_0) & = \gN(\bxti{0}{i}, \sigma_{\rvy}^2 / s_i^2). \label{eq:qyx}
\end{align}

\textbf{Case I } For $\rvx_T$, it is obvious when $s_i = 0$. When $s_i > 0$, we have \autoref{eq:qyx} and that:
\begin{align}
    q^{(T)}(\bxti{T}{i} | \rvx_0, \rvy) &= \gN(\byi{i}, \sigma_{T}^2 - \frac{\sigma_{\rvy}^2}{s_i^2}),
\end{align}
and thus 
\begin{align}
    q^{(T)}(\bxti{T}{i} | \rvx_0) & = \gN(\bxti{0}{i}, \sigma_{\rvy}^2 / s_i^2 + \sigma_{T}^2 - \frac{\sigma_{\rvy}^2}{s_i^2}) = \gN(\bxti{0}{i}, \sigma_{T}^2). \nonumber
\end{align}

\textbf{Case II } For any $t < T$ and $i$ such that $s_i > 0$ and $\sigma_t > \sigma_{\rvy} / s_i$, we have \autoref{eq:qyx} and that:
\begin{align}
    q^{(t)}(\bxti{t}{i} | \rvx_{t+1}, \rvx_0, \rvy) &= \gN\left((1 - \eta_b) \bxti{0}{i} + \eta_b \byi{i}, \sigma_{t}^2 - \frac{\sigma_{\rvy}^2}{s_i^2} \eta_b^2\right),
\end{align}
and thus we can safely remove the dependence on $\rvx_{t+1}$ via marginalization.  $q^{(t)}(\bxti{t}{i} | \rvx_0)$ is a Gaussian with the mean being $(1 - \eta_b) \bxti{0}{i} + \eta_b \bxti{0}{i} = \bxti{0}{i}$ and variance being
$$
\sigma_{t}^2 - \frac{\sigma_{\rvy}^2}{s_i^2} \eta_b^2 + \frac{\sigma_{\rvy}^2}{s_i^2} \eta_b^2 = \sigma_{t}^2,
$$
where we note that $\byi{i}$ has a standard deviation of $\sigma_\rvy / s_i$.

\textbf{Case III } For any $t < T$ and $i$ such that $s_i > 0$ and $\sigma_t < \sigma_{\rvy} / s_i$, we have \autoref{eq:qyx},
 so $(\byi{i} - \bxti{0}{i}) / (\sigma_{\rvy} / s_i)$ is distributed as a standard Gaussian.
 Moreover, similar to \textbf{Case II},
     $q^{(t)}(\bxti{t}{i} | \rvx_0)$ is a Gaussian with its mean being 
    $$\bxti{0}{i} + \sqrt{1 - \eta^2} \sigma_{t} \frac{\byi{i} - \bxti{0}{i}}{\sigma_{\rvy} / s_i}$$ and its variance being $\eta^2 \sigma_{t}^2$, 
so $q^{(t)}(\bxti{t}{i} | \rvx_0)$ is a Gaussian with a mean of $\bxti{0}{i}$ and a variance of
$$
 (1 - \eta^2) \sigma_t^2 + \eta^2 \sigma_t^2 = \sigma_t^2.
$$

\textbf{Case IV } For any $t \leq T$ and $i$ such that $s_i = 0$ (where there is no dependence on $\rvy$), we apply mathematical induction. The base case ($t = T$) is true, as we have shown earlier in \textbf{Case I}. In the step case ($t < T$), we have that $q^{(t+1)}(\bxti{t+1}{i} | \rvx_0) = \gN(\bxti{0}{i}, \sigma_{t+1}^2)$. Similar to \textbf{Case II}, $q^{(t)}(\bxti{t}{i} | \rvx_0)$ is a Gaussian with its mean being
$$
    \bxti{0}{i} + \sqrt{1 - \eta^2} \sigma_{t} \frac{\bxti{t+1}{i} - \bxti{0}{i}}{\sigma_{t+1}}
$$
and variance being $\eta^2 \sigma_{t}^2$, which does not depend on $\rvy$.
Therefore, $q^{(t)}(\bxti{t}{i} | \rvx_0)$ is also Gaussian, with a mean of $\bxti{0}{i}$ and a variance of
$$
(1 - \eta^2) \sigma_{t}^2 + \eta^2 \sigma_{t}^2 = \sigma_t^2.
$$
Hence, the proof is completed via the four cases.
\end{proof}

\ddrmmain*

\begin{proof}

As there is no parameter sharing between models at different time steps $t$, let us focus on any particular time step $t$ and rewrite the corresponding objective as a denoising autoencoder objective.

\textbf{Case I } For $t > 0$, the only term in \autoref{eq:elbo} that is related to $f^{(t)}_\theta$ (which is used to make the prediction $\rvx_{\theta, t}$) is:
\begin{align}
    & \KL(q^{(t)}(\rvx_t | \rvx_{t+1}, \rvx_{0}, \rvy) \Vert p_\theta^{(t)}(\rvx_t | \rvx_{t+1}, \rvy)) \nonumber \\
    = \ & \KL(q^{(t)}(\bar{\rvx}_t | \rvx_{t+1}, \rvx_{0}, \rvy) \Vert p_\theta^{(t)}(\bar{\rvx}_t | \rvx_{t+1}, \rvy)) \nonumber \\
    = \ & \sum_{i=1}^{n} \KL(q^{(t)}(\bxti{t}{i} | {\rvx}_{t+1}, {\rvx}_{0}, \rvy) \Vert p_\theta^{(t)}(\bxti{t}{i} | {\rvx}_{t+1}, {\rvx}_{0}, \rvy)), \label{eq:elbo-decomp}
\end{align}
where the first equality is from the orthogonality of $\mV^\top$ and the second equality is from the fact that both $q^{(t)}$ and $p_\theta^{(t)}$ over the spectral space are Gaussians with identical diagonal covariance matrices (so the KL divergence can factorize).

Here, we will use a simple property of the KL divergence between univariate Gaussians~\cite{kingma2013auto}:
\begin{quote}
If $p = \gN(\mu_1, V_1)$, $q = \gN(\mu_2, V_2)$, then $$\KL(p \Vert q) = \frac{1}{2} \log \frac{V_2}{V_1} + \frac{V_1 + (\mu_1 - \mu_2)^2}{2 V_2} - \frac{1}{2}.$$
\end{quote}

Since we constructed $p_\theta^{(t)}$ and $q^{(t)}$ to have the same variance, \autoref{eq:elbo-decomp} is a total squared error with weights for each dimension of $\bar{\rvx}_t$ (the spectral space), so the DDPM objective (which is a total squared error objective in the original space) is still a good approximation. In order to transform it into a denoising autoencoder objective (equivalent to DDPM), the weights have to be equal. Next, we will show that our construction of $\eta = 1$ and $\eta_b = 2 \sigma_t^2 / (\sigma_t^2 + \sigma_\rvy^2 / s_i^2)$ satisfies this.

All the indices $i$ will fall into one of the three cases: $s_i = 0$, $\sigma_t < \sigma_{\rvy}/s_i$, or $\sigma_t > \sigma_{\rvy}/s_i$. 
\begin{itemize}
    \item For $s_i = 0$, the KL divergence is
$
\frac{(\bxthetai{i} - \bxti{0}{i})^2}{2  \sigma_t^2},
$
where we recall ${\color{violet}\bar{\rvx}_{\theta, t}} = \mV^\top f_\theta^{(t)}(\rvx_{t+1})$.
\item For $\sigma_t < \frac{\sigma_{\rvy}}{s_i}$, the KL divergence is also
$
\frac{(\bxthetai{i} - \bxti{0}{i})^2}{2 \sigma_t^2}.
$
\item For $\sigma_t \geq \frac{\sigma_{\rvy}}{s_i}$, we have defined $\eta_b$ as a solution to the following quadratic equation (the other solution is $0$, which is irrelevant to our case since it does not make use of information from $\rvy$):
\begin{align}
    (\sigma_t^2  + \frac{\sigma_{\rvy}^2}{s_i^2}) \eta_b^2 - 2\sigma_t^2  \eta_b = 0;
\end{align}
reorganizing terms, we have that:
\begin{gather}
    (\sigma_t^2  + \frac{\sigma_{\rvy}^2}{s_i^2}) \eta_b^2 - 2\sigma_t^2  \eta_b + \sigma_t^2 = \sigma_t^2 \nonumber \\
    \sigma_t^2 (1 - \eta_b)^2 = \sigma_t^2 \eta_b^2 - 2\sigma_t^2  \eta_b + \sigma_t^2 = \sigma_t^2 - \frac{\sigma_{\rvy}^2}{s_i^2} \eta_b^2 \nonumber \\
    \frac{(1 - \eta_b)^2}{\sigma_t^2 - \frac{\sigma_{\rvy}^2}{s_i^2} \eta_b^2} = \frac{1}{\sigma_t^2},
\end{gather}
So the KL divergence is
\begin{align*}
    \frac{(1 - \eta_b)^2}{2(\sigma_t^2 - \frac{\sigma_{\rvy}^2}{s_i^2} \eta_b^2)} (\bxthetai{i} - \bxti{0}{i})^2 = \frac{(\bxthetai{i} - \bxti{0}{i})^2}{2 \sigma_t^2}.
\end{align*}
\end{itemize}

Therefore, regardless of how the cases are distributed among indices, we will always have that:
\begin{align*}
    & \KL(q^{(t)}(\bar{\rvx}_t | \rvx_{t+1}, \rvx_{0}, \rvy) \Vert p_\theta^{(t)}(\bar{\rvx}_t | \rvx_{t+1}, \rvy)) =  \sum_{i=1}^{n^2} \frac{(\bxthetai{i} - \bxti{0}{i})^2}{2\sigma_t^2 } = \frac{\norm{{\color{violet}\bar{\rvx}_{\theta, t}} - \bxt{0}}_2^2}{2\sigma_t^2} =  \frac{\norm{f_\theta^{(t)}(\rvx_{t+1}) - \rvx_0}_2^2}{2\sigma_t^2 }.
\end{align*}

\textbf{Case II } For $t = 0$, we will only have two cases ($s_i = 0$ or $\sigma_t < \frac{\sigma_{\rvy}}{s_i}$), and thus, similar to \textbf{Case I},
\begin{align*}
\log p_\theta^{(0)}(\bar{\rvx}_0 | \rvx_{1}, \rvy) = \sum_{i=1}^{n^2} \log p_\theta^{(0)}(\bxti{0}{i} | \rvx_{1}, \rvy)
\propto \sum_{i=1}^{n^2} ({\color{violet}\bar{\rvx}_{\theta, 0}^{(i)}} - \bxti{0}{i})^2 = \norm{{\color{violet}\bar{\rvx}_{\theta, 0}} - \bar{\rvx}_0}_2^2 = \norm{f_\theta^{(0)}(\rvx_{1}) - \rvx_0}_2^2,
\end{align*}
as long as we have a constant variance for $p_\theta^{(0)}$.
Thus, every individual term in \autoref{eq:elbo} can be written as a denoising autoencoder objective, completing the proof.
\end{proof}
\section{Memory Efficient SVD}
\label{sec:svd_expl}
Here we explain how we obtained the singular value decomposition (SVD) for different degradation models efficiently.

\subsection{Denoising}
In denoising, the corrupted image is the original image with additive white Gaussian noise. Therefore, $\mH = \mI$ and all the SVD elements of $\mH$ are simply the identity matrix $\mI$, which in turns makes their multiplication by different vectors trivial.

\subsection{Inpainting}
In inpainting, $\mH$ retains a known subset of size $k$ of the image's pixels. This is equivalent to permuting the pixels such that the retained one are placed at the top, then keeping the first $k$ entries. Therefore,
\begin{align}
\label{eqn:svd_inp}
    \mH = \mI \mSigma \mP,
\end{align}
where $\mP$ is the appropriate permutation matrix, $\mSigma$ is a rectangular diagonal matrix of size $k \times n$ with ones in its main diagonal, and $\mI$ is the identity matrix. Since permutation matrices are orthogonal, \autoref{eqn:svd_inp} is the SVD of $\mH$.

We can multiply a given vector by $\mP$ and $\mP^T$ by storing the permutation itself rather than the matrix. $\mSigma$ can multiply a vector by simply slicing it. Therefore, by storing the appropriate permutation and the number $k$, we can apply each element of the SVD with $\Theta(n)$ space complexity.

\subsection{Super Resolution}
For super resolution, we assume that the original image of size $d \times d$ (\emph{i.e.} $n = 3 d^2$) is downscaled using a block averaging filter by $r$ in each dimension, such that $d$ is divisible by $r$.
In this scenario, each pixel in the output image is the average of an $r \times r$ patch in the input image, and each such patch affects exactly one output pixel.
Therefore, any output pixel is given by 
$(\mH \rvx)_i = \vk^T \vp_i$, where $\vk$ is a vector of size $r^2$ with $\frac{1}{r^2}$ in each entry, and $\vp_i$ is the vectorized $i$-th $r \times r$ patch.
More formally, if $\mP_1$ is a permutation matrix that reorders a vectorized image into patches, then 
\begin{align*}
    \mH = \left( \mI \otimes \vk^T \right) \mP_1,
\end{align*}
where $\otimes$ is the Kronecker product, and $\mI$ is the identity matrix of size $\frac{d}{r} \times \frac{d}{r}$.
In order to obtain the SVD of $\mH$, we calculate the SVD of $\vk^T$:
\begin{align*}
    \vk^T = \mU_\vk \mSigma_\vk \mV_\vk^T.
\end{align*}
Using properties of the Kronecker product, we observe
\begin{align}
    \mH & = \left( \mI \otimes \vk^T \right) \mP_1
     = \left( \left( \mI \mI \mI \right) \otimes \left( \mU_\vk \mSigma_\vk \mV_\vk^T \right) \right) \mP_1 \\
    & = \left( \mI \otimes \mU_\vk \right)  \left( \mI \otimes \mSigma_\vk \right) \left( \mI \otimes \mV_\vk^T \right) \mP_1. \nonumber
\end{align}
The Kronecker product of two orthogonal matrices is an orthogonal matrix.
Therefore, $\mI \otimes \mU_\vk$ and $\mI \otimes \mV_\vk^T$ are orthogonal.
Observe that the matrix $\mI \otimes \mSigma_\vk$ has one non-zero value ($\frac{1}{r^2}$) in each row. By applying a simple permutation on its columns, these values can be reordered to be on the main diagonal. We denote the appropriate permutation matrix by $\mP_2$, and obtain
\begin{align}
\label{eqn:svd_sr}
    \mH = \mU \mSigma \mV^T,
\end{align}
where $\mU = \mI \otimes \mU_\vk$ is orthogonal,
$\mSigma = \left( \mI \otimes \mSigma_\vk \right) \mP_2^T$ is a rectangular diagonal matrix with non-negative entries, and
$\mV^T = \mP_2 \left( \mI \otimes \mV_\vk^T \right) \mP_1$ is orthogonal.
As such, \autoref{eqn:svd_sr} is the SVD of $\mH$.
By storing the permutations and the SVD elements of $\vk^T$, we can simulate each element of the SVD of $\mH$ with $\Theta(n)$ space complexity, without directly calculating the Kronecker products with $\mI$.

\subsection{Colorization}
The grayscale image is obtained by averaging the red, green, and blue channels of each pixel.
This means that every output pixel is given by $\left(\mH \rvx\right)_i = \vk^T \vp_i$, where $\vk^T = \begin{pmatrix} \frac{1}{3} & \frac{1}{3} & \frac{1}{3} \end{pmatrix}$ and $\vp_i$ is the $3$-valued $i$-th pixel of the original color image.
The SVD of $\mH$ is obtained exactly the same as in the super resolution case, with separate pixels replacing separate patches.

\subsection{Deblurring}
We focus on \textit{separable blurring}, where the $2$D blurring kernel is $\mK = \vr \vc^T$, which means $\vc$ is applied on the columns of the image, and $\vr^T$ is applied on its rows.
The blurred image can be obtained by $\mB = \mA_c \mX \mA_r^T$, where $\mA_c$ and $\mA_r$ apply a $1$D convolution with kernels $\vc$ and $\vr$, respectively.
Alternatively, $\vb = \mH \vx$, where $\vx$ is the vectorized image $\mX$, $\vb$ is the vectorized blurred image $\mB$, and $\mH$ is the matrix applying the $2$D convolution $\mK$.
It can be shown that $\mH = \mA_r \otimes \mA_c$, where $\otimes$ is the Kronecker product.
In order to calculate the SVD of $\mH$, we calculate the SVD of $\mA_r$ and $\mA_c$:
\begin{align*}
    \mA_r = \mU_r \mSigma_r \mV_r^T, \quad
    \mA_c = \mU_c \mSigma_c \mV_c^T.
\end{align*}
Using the properties of the Kronecker product, we observe
\begin{equation}
\label{eqn:svd_deblur_almost}
\begin{aligned}
    \mH & = \mA_r \otimes \mA_c =
    \left(\mU_r \mSigma_r \mV_r^T\right) \otimes \left(\mU_c \mSigma_c \mV_c^T\right) \\
    & = \left(\mU_r \otimes \mU_c\right) \left(\mSigma_r \otimes \mSigma_c\right) \left(\mV_r \otimes \mV_c\right)^T.
\end{aligned}
\end{equation}
The Kronecker product preserves orthogonality. Therefore, \autoref{eqn:svd_deblur_almost} is a valid SVD of $\mH$, with the exception of the singular values not being on the main diagonal, and not being sorted descendingly.
We reorder the columns so that the singular values are on the main diagonal and denote the corresponding permutation matrix by $\mP_1$.
We also sort the values descendingly and denote the sorting permutation matrix by $\mP_2$, and obtain the following SVD:
\begin{align}
\label{eqn:svd_deblur}
    \mH = \mU \mSigma \mV^T,
\end{align}
where $\mU = \left(\mU_r \otimes \mU_c\right) \mP_2^T$, $\mSigma = \mP_2 \left(\mSigma_r \otimes \mSigma_c\right) \mP_1^T \mP_2^T$, and $\mV^T = \mP_2 \mP_1 \left(\mV_r \otimes \mV_c\right)^T$.

For every matrix of the form $\mM = \mN \otimes \mL$, it holds that $\mM x$ is the vectorized version of $\mL \mX \mN^T$. By using this property and applying the relevant permutation, we can simulate multiplying a vector by $\mU$, $\mV$, $\mU^T$, or $\mV^T$ without storing the full matrix.
The space complexity of this approach is $\Theta(n)$, which is required for computing the SVD of $\mA_r$ and $\mA_c$, as well as storing the permutations.

The above calculations remain valid when the blurring is zero-padded, \emph{i.e.}, images are padded with zeroes so that the convolution is not circulant around the edges. We consider a zero-padded deblurring problem in our experiments.
Note that the noiseless version of this problem has a simple solution -- applying the pseudo-inverse of the blurring matrix on the blurry image. This solution attains $32.41$dB in PSNR on ImageNet-1K, while DDRM improves upon it and achieves $35.64$dB.
When noise is added to the blurry image, such a simple solution amplifies the noise and fails to provide a valid output. Therefore, we opt not to report its results.

Furthermore, the above calculations are also applicable to blurring with strided convolutions. We use this fact in our implementation of the bicubic super resolution SVD, which can be interpreted as a strided convolution with a fixed kernel.


\section{Ablation Studies on Hyperparameters}
\label{sec:eta_ablation}

\paragraph{$\mathbf{\eta}$ and $\mathbf{\eta_b}$.}
Apart from the timestep schedules, DDRM has two hyperparameters $\eta$ and $\eta_b$, which control the level of noise injected at each timestep. To identify an ideal combination, we perform a hyperparameter search over $\eta, \eta_b \in \{0.7, 0.8, 0.9, 1.0\}$ for the task of deblurring with $\sigma_y = 0.05$ in $1000$ ImageNet validation images, using the model trained in~\cite{guided_diffusion}. It is possible to also consider different $\eta$ values for $s_i = 0$ and $\sigma_i < \sigma_\rvy / s_i$; we leave that as future work.

We report PSNR and KID results in \autoref{tab:app:ablation}. 
From the results, we observe that generally (\textit{i}) as $\eta_b$ increases, PSNR increases while KID decreases, which is reasonable given that we wish to leverage the information from $\rvy$; (\textit{ii}) as $\eta$ increases, PSNR increases (except for $\eta = 1.0$) yet KID also increases, which presents a trade-off in reconstruction error and image quality (known as the perception-distortion trade-off~\cite{blau2018perception}). Therefore, we choose $\eta_b = 1$ and $\eta = 0.85$ to balance performance on PSNR and KID when we report results.

\paragraph{Timestep schedules.} The timestep schedule has a direct impact on NFEs, as the wall-clock time is roughly linear with respect to NFEs~\cite{song2020denoising}. In Tables~\ref{tab:imagenet_sr4} and~\ref{tab:imagenet_deblur_noisy}, we compare the PSNR, FID, and KID of DDRM with 20 or 100 timesteps (with or without conditioning) and default $\eta = 0.85$ and $\eta_b = 1$. We observe that DDRM with 20 or 100 timesteps have similar performance when other hyperparameters are identical, with DDRM (20) having a slight edge in FID and KID.

\begin{table*}
\centering
\caption{Ablation studies on $\eta$ and $\eta_b$.}
\label{tab:app:ablation}
\begin{subtable}{0.45\textwidth}
\centering
\caption{PSNR $(\uparrow)$.}
\begin{tabular}{l|cccc}
\toprule
    \backslashbox{$\eta$}{$\eta_b$} & 0.7 & 0.8 & 0.9 & 1.0 \\\midrule
     0.7 & 25.16 & 25.19 & 25.20 & 25.20 \\
     0.8 & 25.17 & 25.23 & 25.27 & 25.29 \\
     0.9 & 25.07 & 25.18 & 25.26 & 25.32 \\
     1.0 & 24.54 & 25.75 & 24.91 & 25.04 \\\bottomrule
\end{tabular}
\end{subtable}
\begin{subtable}{0.45\textwidth}
\centering
\caption{KID $\times 10^3$  $(\downarrow)$.}
\begin{tabular}{l|cccc}
\toprule
    \backslashbox{$\eta$}{$\eta_b$} & 0.7 & 0.8 & 0.9 & 1.0 \\\midrule
     0.7 & 16.27 & 14.30 & 12.76 & 11.65 \\
     0.8 & 21.07 & 19.07 & 17.37 & 15.98 \\
     0.9 & 27.85 & 25.64 & 23.81 & 22.40 \\
     1.0 & 45.10 & 42.50 & 40.10 & 37.84 \\\bottomrule
\end{tabular}
\end{subtable}
\end{table*}

\section{Experimental Setup of DGP, RED, and SNIPS}
\label{sec:comp_setup}

Recall that we evaluated DGP~\cite{dgp}, RED~\cite{red}, and SNIPS~\cite{snips} on $256\times256$ ImageNet 1K images, for the problems of $4\times$ super resolution and deblurring without any noise in the measurements. Below we expand on the experimental setup of each one.

For DGP~\cite{dgp}, we use the same hyperparameters introduced in the original paper for MSE-biased super resolution.
We note that the downscaling applied in DGP is different from the block averaging filter that we used, and the numbers they reported are on the $128\times128$ resolution.
Nevertheless, in our experiments, DGP achieved a PSNR of $23.06$ on ImageNet 1K $256\times256$ block averaging $4\times$ super resolution, which is similar to the $23.30$ reported in the original work.
When applied on the deblurring problem, we retained the same DGP hyperparameters as well.

For RED~\cite{red}, we apply the iterative algorithm only in the luminance channel of the image in the YCbCr space, as done in the original paper for deblurring and super resolution.
As for the denoising engine enabling the algorithm, we use the same diffusion model used in DDRM to enable as fair a comparison as possible. We use the last step of the diffusion model (equivalent to denoising with $\sigma = 0.005$), as we found it to work best empirically.
We also chose the steepest-descent version (RED-SD), and $\lambda = 500$ for best PSNR performance given the denoiser we used. We also set $\sigma_0 = 0.01$ when the measurements are noiseless, because $\sigma_0$ cannot be $0$ as RED divides by it.

In super resolution, RED is initialized with the bicubic upsampled low-res image. In deblurring, it is initialized with the blurry image. We then run RED on the ImageNet 1K for different numbers of steps (see \autoref{tab:red_app}), and choose the best PSNR for each problem. Namely, we show in our paper RED on super resolution with $100$ steps, and on deblurring with $500$ steps.
Interestingly, RED achieves a PSNR close to its best for super resolution in just $20$ steps. However, DDRM (with $20$ steps) still outperforms RED in PSNR, with substantially better perceptual quality (see Table 1 in the main paper).

Another interesting plug-and-play image restoration method is DPIR~\cite{dpir}, which has recently achieved impressive results. It does so by applying the well-known Half Quadratic Splitting (HQS) plug-and-play algorithm using a newly proposed architecture.
HQS requires an analytical solution of a minimization problem which is infeasible in general, due to the high memory requirements. DPIR provides efficient solutions for the specific degradation matrices $\mathbf{H}$ considered (circulant blurring, bicubic downsampling), which are different from the ones we consider (zero-padded blurring, block downsampling).
In order to draw a fair comparison between the algorithms, one would have to use the same denosier architecture in both (as we have done for RED and SNIPS), and use the same degradation models.
To apply DPIR on the same problems that we consider, we would need to substantially modify it and introduce efficient solutions. Therefore, we instead compare to RED, an alternative plug-and-play method.

\begin{table}
    \centering
    \caption{RED results on ImageNet 1K ($256\times256$) for $4\times$ super resolution and deblurring for different numbers of steps.}
    \label{tab:red_app}
    \vskip 0.15in
    \begin{center}
    \begin{small}
    \begin{sc}
    \begin{tabular}{lcccc}
        \toprule
        & \multicolumn{2}{c}{Super-res}
        & \multicolumn{2}{c}{Deblurring} \\
        Steps & PSNR$\uparrow$ & KID$\downarrow$ & PSNR$\uparrow$ & KID$\downarrow$ \\
        \midrule
        $0$ & $25.65$ & $44.90$ & $19.26$ & $38.00$ \\
        $20$ & $26.05$ & $52.51$ & $23.49$ & $21.99$ \\
        $100$ & $26.08$ & $53.55$ & $25.00$ & $26.09$ \\
        $500$ & $26.00$ & $54.19$ & $26.16$ & $21.21$ \\
        \bottomrule
    \end{tabular}
    \end{sc}
    \end{small}
    \end{center}
    \vskip -0.1in
\end{table}

SNIPS~\cite{snips} did not originally work with ImageNet images. However, considering the method's similarity to DDRM (as both operate in the spectral space of $\mH$), a comparison is necessary.
We apply SNIPS with the same underlying diffusion model (with all $1000$ timesteps) as DDRM for fairness.
SNIPS evaluates the diffusion model $\tau$ times for each timestep. We set $\tau = 1$ so that SNIPS' runtime remains reasonable in comparison to the rest of the considered methods, and do not explore higher values of $\tau$. It is worth mentioning that in the original work, $\tau$ was set to $3$ for an LSUN bedrooms diffusion model with $1086$ timesteps. We set $c = 0.67$ as it achieved the best PSNR performance.

The original work in SNIPS calculates the SVD of $\mH$ directly, which hinders its ability to handle $256 \times 256$ images on typical hardware. In order to draw comparisons, we replaced the direct calculation of the SVD with our efficient implementation detailed in \autoref{sec:svd_expl}.

In Figure 4 and Table 2 in the main paper, we show that DGP, RED, and SNIPS all fail to produce viable results when significant noise is added to the measurements.
For these results, we use the same hyperparameters used in the noiseless case for all algorithms (except $\sigma_\rvy$ where applicable). While tuning the hyperparameters may boost performance, we do not explore that option as we are only interested in algorithms where given $\mH$ and $\sigma_\rvy$, the restoration process is automatic.
To further demonstrate DDRM's capabilities and speed, we evaluate it on the entire $50,000$-image ImageNet validation set in Tables \ref{tab:imagenet_sr4} and \ref{tab:imagenet_deblur_noisy}, reporting Fréchet Inception distance (FID) \cite{fid} as well as KID, as enough samples are available.

\begin{table}
    \centering
    \caption{ImageNet 50K validation set ($256\times256$) results on $4\times$ super resolution with additive noise of $\sigma_{\rvy} = 0.05$.}
    \label{tab:imagenet_sr4}
    \vskip 0.15in
    \begin{center}
    \begin{small}
    \begin{sc}
    \begin{tabular}{lcccc}
        \toprule
        Method & PSNR$\uparrow$ & FID$\downarrow$ & KID$\downarrow$ & NFEs$\downarrow$ \\
        \midrule
        Bicubic & $22.65$ & $64.24$ & $50.56$ & \first{0} \\
        DDRM & \second{24.70} & $20.16$ & $15.25$ & $100$ \\
        DDRM-CC & \first{24.71} & $18.22$ & $13.57$ & $100$ \\
        DDRM & $24.29$ & \second{17.88} & \second{13.18} & \second{20} \\
        DDRM-CC & $24.30$ & \first{15.92} & \first{11.47} & \second{20} \\
        \bottomrule
    \end{tabular}
    \end{sc}
    \end{small}
    \end{center}
    \vskip -0.1in
\end{table}
\begin{table}
    \centering
    \caption{ImageNet 50K validation set ($256\times256$) results on deblurring with additive noise of $\sigma_{\rvy} = 0.05$.
    }
    \label{tab:imagenet_deblur_noisy}
    \vskip 0.15in
    \begin{center}
    \begin{small}
    \begin{sc}
    \begin{tabular}{lcccc}
        \toprule
        Method & PSNR$\uparrow$ & FID$\downarrow$ & KID$\downarrow$ & NFEs$\downarrow$ \\
        \midrule
        Blurry & $18.05$ & $93.36$ & $74.13$ & \first{0} \\
        DDRM & $24.23$ & $22.30$ & $16.23$ & $100$ \\
        DDRM-CC & $24.21$ & \second{20.06} & \second{14.20} & $100$ \\
        DDRM & \second{24.60} & $21.60$ & $15.65$ & \second{20} \\
        DDRM-CC & \first{24.61} & \first{19.66} & \first{13.94} & \second{20} \\
        \bottomrule
    \end{tabular}
    \end{sc}
    \end{small}
    \end{center}
    \vskip -0.1in
\end{table}

\section{Runtime of Algorithms} 
\label{sec:runtime}
In the main paper, we show the number of neural function evaluations (NFEs) as a proxy for the runtime of algorithms.
Here, we consider the case of noisy deblurring, and measure the runtime of DDRM, RED, SNIPS, and DGP on an Nvidia RTX 3080 GPU.
For each image, DDRM, RED, and SNIPS all run at around $0.09$ s/it (seconds per iteration), with negligible differences of $<0.01$s/it.
We note that the denoiser model of DDRM, SNIPS, and RED is the same, so runtime is almost perfectly linearly correlated with NFEs.
As for DGP, it uses a different model (a GAN), and it is slightly slower than our denoiser ($0.11$ s/it); this is partly because DGP requires additional gradient computations in order to perform an update.
All in all, we observe that the runtime is indeed linear with NFEs, and since no algorithm has a significant runtime advantage over the rest, we prefer to use NFEs as a proxy for runtime, as it is a hardware-independent measure.

In this paper, we used pretrained generative models for image restoration. Since we didn't train any models, a single Nvidia RTX 3080 GPU was sufficient to run all experiments that were shown in the paper and the appendices.

\section{ILVR as a special case of DDRM}
\label{sec:ilvr_proof}

Given a generative diffusion model (\emph{e.g.} DDPM~\cite{ddpm}) that can predict $\rvx$ given $\rvx_{t+1}$ and $t+1$ for $t \in [0, T-1]$, and a noiseless measurement $\rvy = \mH \rvx$, where $\mH$ is a downscaling matrix, the Iterative Latent Variable Refinement (ILVR) \cite{ilvr} algorithm can sample from the posterior distribution $p_\theta^{(t)}(\rvx_{t} | \rvx_{t+1}, \rvy)$ for $t \in [0, T-1]$.

We assume a variance exploding diffusion model, \emph{i.e.} $\rvx_t = \rvx + \sigma_t \epsilon_t$ where $\epsilon_t \sim \gN(0, \mI)$, without loss of generality (because it is equivalent to the variance preserving scheme, as we show in~\autoref{app:sec:ve-vp}). Under this setting, ILVR applies the following updates for $t = T-1, \dots, 0$:
\begin{align*}
    \rvx_{t}^{\prime} & = \textcolor{violet}{\rvx_{\theta, t}(\rvx_{t+1}, t+1)} + \sigma_t \epsilon_t, \\
    \rvy_{t} & = \mH^{\dagger} \rvy + \sigma_t \epsilon_t^{\prime}, \\
    \rvx_{t} & = \rvx_{t}^{\prime} - \mH^{\dagger} \mH \rvx_{t}^{\prime} + \mH^{\dagger} \mH \rvy_{t},
\end{align*}
where $\textcolor{violet} {\rvx_{\theta, t}(\rvx_{t+1}, t+1)}$ is the prediction for $\rvx$ given by the diffusion model at timestep $t+1$, $\epsilon_t \sim \gN(0, \mI)$, and $\epsilon_t^{\prime} \sim \gN(0, \mI)$.
Substituting $\rvx_{t}^{\prime}$, $\rvy_{t}$, and $\mH = \mU \mSigma \mV^T$, the last equation becomes
\begin{align*}
    \rvx_{t} & = \rvx_{t}^{\prime} - \mH^{\dagger} \mH \rvx_{t}^{\prime} + \mH^{\dagger} \mH \rvy_{t} \\
    & = \textcolor{violet}{\rvx_{\theta, t}(\rvx_{t+1}, t+1)} + \sigma_t \epsilon_t - \mH^{\dagger} \mH \left( \textcolor{violet}{\rvx_{\theta, t}(\rvx_{t+1}, t+1)} + \sigma_t \epsilon_t \right) + \mH^{\dagger} \mH \left( \mH^{\dagger} \rvy + \sigma_t \epsilon_t^{\prime} \right) \\
    & = \textcolor{violet}{\rvx_{\theta, t}(\rvx_{t+1}, t+1)} + \sigma_t \epsilon_t - \mV \mSigma^{\dagger} \mU^T \mU \mSigma \mV^T \left( \textcolor{violet}{\rvx_{\theta, t}(\rvx_{t+1}, t+1)} + \sigma_t \epsilon_t \right) + \mV \mSigma^{\dagger} \mU^T \mU \mSigma \mV^T \left( \mV \mSigma^{\dagger} \mU^T \rvy + \sigma_t \epsilon_t^{\prime} \right) \\
    & = \textcolor{violet}{\rvx_{\theta, t}(\rvx_{t+1}, t+1)} + \sigma_t \epsilon_t - \mV \mSigma^{\dagger} \mSigma \mV^T \left( \textcolor{violet}{\rvx_{\theta, t}(\rvx_{t+1}, t+1)} + \sigma_t \epsilon_t \right) + \mV \mSigma^{\dagger} \mSigma \mV^T \left( \mV \mSigma^{\dagger} \mU^T \rvy + \sigma_t \epsilon_t^{\prime} \right) \\
    & = \textcolor{violet}{\rvx_{\theta, t}(\rvx_{t+1}, t+1)} + \sigma_t \epsilon_t - \mSigma^{\dagger} \mSigma \mV \mV^T \left( \textcolor{violet}{\rvx_{\theta, t}(\rvx_{t+1}, t+1)} + \sigma_t \epsilon_t \right) + \mSigma^{\dagger} \mSigma \mV \mV^T \left( \mV \mSigma^{\dagger} \mU^T \rvy + \sigma_t \epsilon_t^{\prime} \right) \\
    & = \textcolor{violet}{\rvx_{\theta, t}(\rvx_{t+1}, t+1)} + \sigma_t \epsilon_t - \mSigma^{\dagger} \mSigma \left( \textcolor{violet}{\rvx_{\theta, t}(\rvx_{t+1}, t+1)} + \sigma_t \epsilon_t \right) + \mSigma^{\dagger} \mSigma \left( \mV \mSigma^{\dagger} \mU^T \rvy + \sigma_t \epsilon_t^{\prime} \right).
\end{align*}
The second to last equality holds because $\mSigma^{\dagger} \mSigma$ is a square diagonal matrix, and matrix multiplication with a square diagonal matrix is commutative. Recall that $\bxt{t} = \mV^T \rvx_{t}$, $\bar{\rvy} = \mSigma^{\dagger} \mU^T \rvy$,  and $\textcolor{violet}{\bar{\rvx}_{\theta,t}} = \mV^T \textcolor{violet}{\rvx_{\theta, t}(\rvx_{t+1}, t+1)}$, thus
\begin{align*}
    \bxt{t} 
    & = \mV^T \textcolor{violet}{\rvx_{\theta, t}(\rvx_{t+1}, t+1)} + \sigma_t \mV^T \epsilon_t - \mV^T \mSigma^{\dagger} \mSigma \left( \textcolor{violet}{\rvx_{\theta, t}(\rvx_{t+1}, t+1)} + \sigma_t \epsilon_t \right) + \mV^T \mSigma^{\dagger} \mSigma \left( \mV \mSigma^{\dagger} \mU^T \rvy + \sigma_t \epsilon_t^{\prime} \right) \\
    & = \mV^T \textcolor{violet}{\rvx_{\theta, t}(\rvx_{t+1}, t+1)} + \sigma_t \mV^T \epsilon_t - \mSigma^{\dagger} \mSigma \mV^T \left( \textcolor{violet}{\rvx_{\theta, t}(\rvx_{t+1}, t+1)} + \sigma_t \epsilon_t \right) + \mSigma^{\dagger} \mSigma \mV^T \left( \mV \mSigma^{\dagger} \mU^T \rvy + \sigma_t \epsilon_t^{\prime} \right) \\
    & = \mV^T \textcolor{violet}{\rvx_{\theta, t}(\rvx_{t+1}, t+1)} + \sigma_t \mV^T \epsilon_t
    - \mSigma^{\dagger} \mSigma \mV^T \textcolor{violet}{\rvx_{\theta, t}(\rvx_{t+1}, t+1)} - \sigma_t \mSigma^{\dagger} \mSigma \mV^T \epsilon_t
    + \mSigma^{\dagger} \mSigma \mV^T \mV \mSigma^{\dagger} \mU^T \rvy + \sigma_t \mSigma^{\dagger} \mSigma \mV^T \epsilon_t^{\prime} \\
    & = \textcolor{violet}{\bar{\rvx}_{\theta,t}} + \sigma_t \mV^T \epsilon_t
    - \mSigma^{\dagger} \mSigma \textcolor{violet}{\bar{\rvx}_{\theta,t}} - \sigma_t \mSigma^{\dagger} \mSigma \mV^T \epsilon_t
    + \mSigma^{\dagger} \mSigma \mSigma^{\dagger} \mU^T \rvy + \sigma_t \mSigma^{\dagger} \mSigma \mV^T \epsilon_t^{\prime} \\
    & = \left( \mI - \mSigma^{\dagger} \mSigma \right) \textcolor{violet}{\bar{\rvx}_{\theta,t}}
    + \left( \mI - \mSigma^{\dagger} \mSigma \right) \sigma_t \mV^T \epsilon_t
    + \mSigma^{\dagger} \mU^T \rvy + \mSigma^{\dagger} \mSigma \sigma_t \mV^T \epsilon_t^{\prime} \\
    & = \left( \mI - \mSigma^{\dagger} \mSigma \right) \textcolor{violet}{\bar{\rvx}_{\theta,t}}
    + \left( \mI - \mSigma^{\dagger} \mSigma \right) \sigma_t \mV^T \epsilon_t
    + \bar{\rvy} + \mSigma^{\dagger} \mSigma \sigma_t \mV^T \epsilon_t^{\prime}.
\end{align*}
The matrix $\mSigma^{\dagger} \mSigma$ is a square diagonal matrix with zeroes in its entries where the singular value is zero, and ones otherwise. In addition, $\mSigma^\dagger$ has a row of zeroes when the singular value is zero.
Therefore, it holds that
\begin{align}
    \bxti{t}{i} & = \begin{cases}
    \textcolor{violet}{\bar{\rvx}_{\theta,t}^{(i)}} + \left( \sigma_t \mV^T \epsilon_t \right)^{(i)} & \text{if } s_i = 0 \\
    \bar{\rvy}^{(i)} + \left( \sigma_t \mV^T \epsilon_t^{\prime}\right)^{(i)} & \text{if } s_i \neq 0
    \end{cases},
\end{align}
which in turn implies
\begin{align}
    p_\theta^{(t)}(\bxti{t}{i} | \rvx_{t+1}, y) & = \begin{cases}
    \gN \left( \textcolor{violet}{\bar{\rvx}_{\theta,t}^{(i)}}, \sigma_t^2 \mI \right) & \text{if } s_i = 0 \\
    \gN \left( \bar{\rvy}^{(i)}, \sigma_t^2 \mI \right) & \text{if } s_i \neq 0
    \end{cases}.
\end{align}
This distribution is exactly the same as Equation 8 in the main paper when $\eta = \eta_b = 1$ and $\sigma_\rvy = 0$.

As for $\rvx_T$, ILVR initializes it by sampling from $\gN\left(0, \sigma_T^2 \mI\right)$ (or $\gN\left(0, \mI\right)$ in the variance preserving case) while DDRM samples according to Equation 7 in the main paper.
The two initializations have the same variance but differ in the mean.
This difference has a negligible effect on the end result since the variance is much larger than the difference in the means.
Therefore, the above form of ILVR is a specific form of a DDRM (with $\eta = \eta_b = 1$), posed as a solution for linear inverse problems without noise in the measurements.

In their experiments, ILVR only tested $\mH$ which is the bicubic downscaling matrix with varying scale factors. In theory, ILVR can also work for any linear degradation $\mH$, as long as $\rvy$ does not contain noise.
\section{Additional Results}
\label{sec:more_pics}
We provide additional figures below showing DDRM's versatility across different datasets, inverse problems, and noise levels (Figures \ref{fig:apdx_faces_sr}, \ref{fig:appendix_faces_other}, \ref{fig:apdx_bedroom}, \ref{fig:apdx_cat}, and \ref{fig:apdx_ood}). We also showcase the sample diversity provided by DDRM in \autoref{fig:apdx_div};
we present more uncurated samples from the ImageNet experiments in Figures \ref{fig:apdx_in1k_sr4} and \ref{fig:apdx_in1k_deblur}.
Moreover, we further illustrate DDRM's advantage over previous unsupervised methods by evaluating on two additional inverse problems:
(i) $4\times$ super-resolution with the popular bicubic downsampling kernel; and
(ii) deblurring with an anisotropic Gaussian blur kernel (with $\sigma=20$ horizontally and $\sigma=1$ vertically), mimicing motion blur.
We show both noiseless and noisy versions in Tables \ref{tab:extra_noiseless} and \ref{tab:extra_noisy}, respectively.
To maintain the unsupervised nature of the tested methods, we use the same hyperparameters as in block-averaging super-resolution and uniform deblurring.

\newpage
\begin{table}
    \centering
    \caption{Noiseless $4\times$ super-resolution (using a bicubic kernel) and anisotropic Gaussian deblurring results on ImageNet 1K ($256\times256$).}
    \label{tab:extra_noiseless}
    \begin{center}
    \begin{tabular}{l p{0.0001\textwidth} cccc p{0.0001\textwidth} cccc}
        \toprule
        \multirow{2}{*}{Method} & ~ &  \multicolumn{4}{c}{$4\times$ super-resolution (Bicubic)} & ~ & \multicolumn{4}{c}{Deblurring (Anisotropic)} \\
        & ~ & PSNR$\uparrow$ & SSIM$\uparrow$ & KID$\downarrow$ & NFEs$\downarrow$ & ~ & PSNR$\uparrow$ & SSIM$\uparrow$ & KID$\downarrow$ & NFEs$\downarrow$ \\
        \midrule
        Baseline & ~ & $26.06$ & \second{0.73} & $72.41$ & \first{0} & ~ & $19.96$ & $0.58$ & $25.23$ & \first{0} \\
        DGP & ~ & $20.82$ & $0.50$ & \second{29.62} & $1500$ & ~ & $23.35$ & $0.59$ & $20.10$ & $1500$ \\
        RED & ~ & \second{26.14} & \second{0.73} & $47.61$ & $100$ & ~ & $29.39$ & \second{0.86} & $10.49$ & $500$ \\
        SNIPS & ~ & $17.65$ & $0.23$ & $30.30$ & $1000$ & ~ & \second{33.34} & \second{0.86} & \second{0.58} & $1000$ \\
        \midrule
        DDRM & ~ & \first{27.09} & \first{0.76} & \first{12.78} & \second{20} & ~ & \first{36.02} & \first{0.93} & \first{0.41} & \second{20} \\ 
        \bottomrule
    \end{tabular}
    \end{center}
\end{table}

\begin{table}
    \centering
    \caption{Noisy ($\sigma_\rvy = 0.05$) $4\times$ super-resolution (using a bicubic kernel) and anisotropic Gaussian deblurring results on ImageNet 1K ($256\times256$).}
    \label{tab:extra_noisy}
    \begin{center}
    \begin{tabular}{l p{0.0001\textwidth} cccc p{0.0001\textwidth} cccc}
        \toprule
        \multirow{2}{*}{Method} & ~ &  \multicolumn{4}{c}{$4\times$ super-resolution (Bicubic)} & ~ & \multicolumn{4}{c}{Deblurring (Anisotropic)} \\
        & ~ & PSNR$\uparrow$ & SSIM$\uparrow$ & KID$\downarrow$ & NFEs$\downarrow$ & ~ & PSNR$\uparrow$ & SSIM$\uparrow$ & KID$\downarrow$ & NFEs$\downarrow$ \\
        \midrule
        Baseline & ~ & $21.68$ & $0.40$ & $73.87$ & \first{0} & ~ & $19.96$ & $0.27$ & $55.00$ & \first{0} \\
        DGP & ~ & $19.68$ & $0.40$ & \second{44.07} & $1500$ & ~ & \second{22.64} & \second{0.53} & \second{25.38} & $1500$ \\
        RED & ~ & \second{22.65} & \second{0.46} & $54.90$ & $100$ & ~ & $11.97$ & $0.10$ & $130.30$ & $500$ \\
        SNIPS & ~ & $16.16$ & $0.14$ & $69.69$ & $1000$ & ~ & $17.49$ & $0.20$ & $48.37$ & $1000$ \\
        \midrule
        DDRM & ~ & \first{25.53} & \first{0.68} & \first{14.57} & \second{20} & ~ & \first{26.95} & \first{0.73} & \first{10.34} & \second{20} \\ 
        \bottomrule
    \end{tabular}
    \end{center}
\end{table}

\renewcommand{\imwidth}{1.25cm}
\begin{figure}[b]
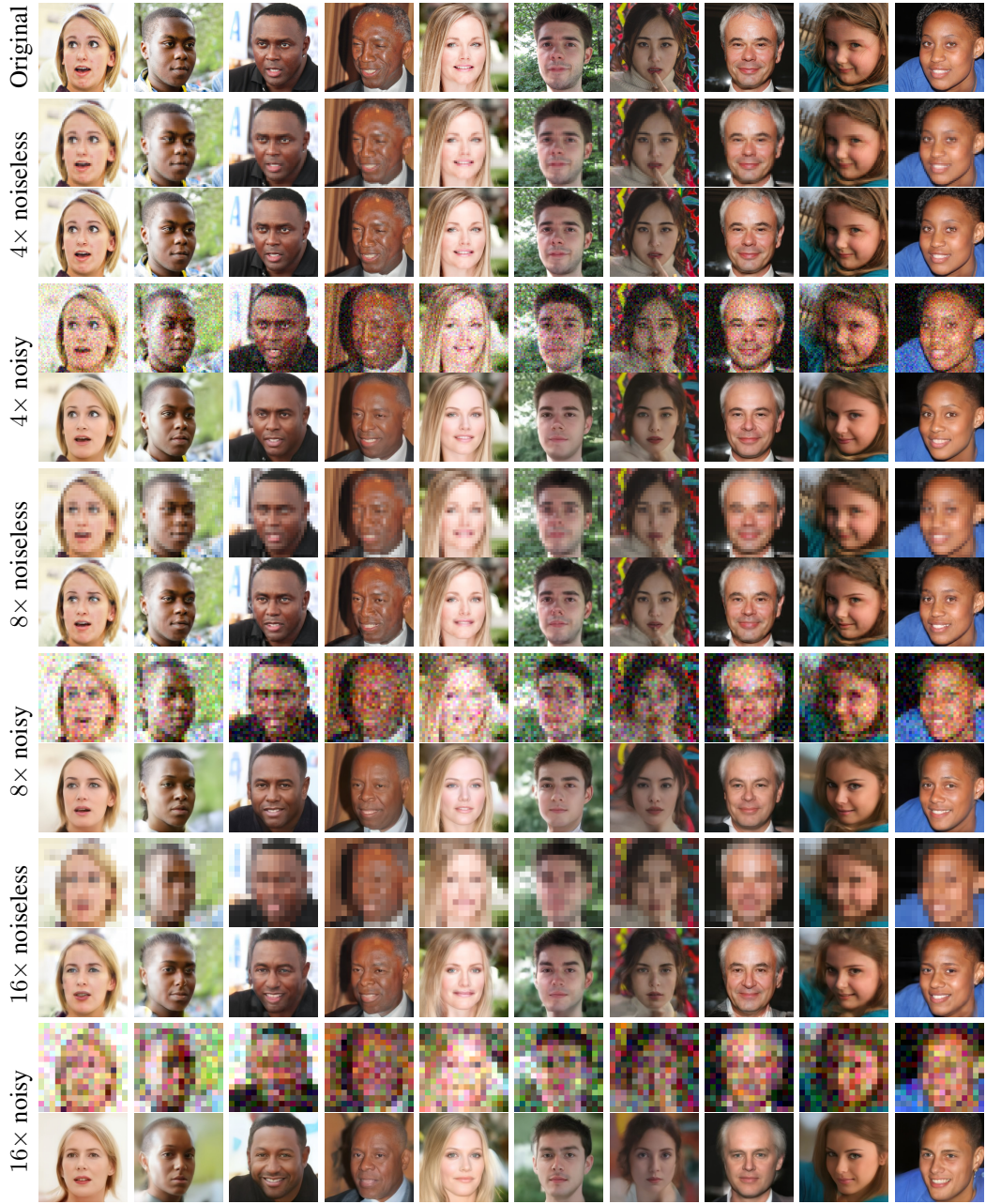

    \centering
    \def\arraystretch{0.2}
    \setlength\tabcolsep{0.005cm}
    \begin{tabular}{l cccccccccc l}
    \raisebox{0.55cm}[0pt][0pt]{\rotatebox[origin=c]{90}{Original}} &
        \forloop{row}{1}{\value{row} < 11}{
            \includegraphics[width=\imwidth,height=\imwidth]{./figures/appendix_faces_sr/sr4_noiseless/orig_\arabic{row}.jpg}
            &
        } \\
        \multicolumn{12}{c}{~} \\
        \multirow{2}{*}{\raisebox{0cm}[0pt][0pt]{\rotatebox[origin=c]{90}{$4\times$ noiseless}}}
        & \forloop{row}{1}{\value{row} < 11}{
            \includegraphics[width=\imwidth,height=\imwidth]{./figures/appendix_faces_sr/sr4_noiseless/y0_\arabic{row}.jpg}
            &
        } \\ 
        & \forloop{row}{1}{\value{row} < 11}{ \includegraphics[width=\imwidth,height=\imwidth]{./figures/appendix_faces_sr/sr4_noiseless/\arabic{row}_s0_-1.jpg}
            &
        } \\ 
        \multicolumn{12}{c}{~} \\
        \multirow{2}{*}{\raisebox{0cm}[0pt][0pt]{\rotatebox[origin=c]{90}{$4\times$ noisy}}}
        & \forloop{row}{1}{\value{row} < 11}{ \includegraphics[width=\imwidth,height=\imwidth]{./figures/appendix_faces_sr/sr4_noisy/y0_\arabic{row}.jpg}
            &
        } \\ 
        & \forloop{row}{1}{\value{row} < 11}{ \includegraphics[width=\imwidth,height=\imwidth]{./figures/appendix_faces_sr/sr4_noisy/\arabic{row}_s0_-1.jpg}
            &
        } \\ 
        \multicolumn{12}{c}{~} \\
        \multirow{2}{*}{\raisebox{0cm}[0pt][0pt]{\rotatebox[origin=c]{90}{$8\times$ noiseless}}}
        & \forloop{row}{1}{\value{row} < 11}{ \includegraphics[width=\imwidth,height=\imwidth]{./figures/appendix_faces_sr/sr8_noiseless/y0_\arabic{row}.jpg}
            &
        } \\ 
        & \forloop{row}{1}{\value{row} < 11}{ \includegraphics[width=\imwidth,height=\imwidth]{./figures/appendix_faces_sr/sr8_noiseless/\arabic{row}_s0_-1.jpg}
            &
        } \\ 
        \multicolumn{12}{c}{~} \\
        \multirow{2}{*}{\raisebox{0cm}[0pt][0pt]{\rotatebox[origin=c]{90}{$8\times$ noisy}}}
        & \forloop{row}{1}{\value{row} < 11}{ \includegraphics[width=\imwidth,height=\imwidth]{./figures/appendix_faces_sr/sr8_noisy/y0_\arabic{row}.jpg}
            &
        } \\ 
        & \forloop{row}{1}{\value{row} < 11}{ \includegraphics[width=\imwidth,height=\imwidth]{./figures/appendix_faces_sr/sr8_noisy/\arabic{row}_s0_-1.jpg}
            &
        } \\ 
        \multicolumn{12}{c}{~} \\
        \multirow{2}{*}{\raisebox{0cm}[0pt][0pt]{\rotatebox[origin=c]{90}{$16\times$ noiseless}}}
        & \forloop{row}{1}{\value{row} < 11}{ \includegraphics[width=\imwidth,height=\imwidth]{./figures/appendix_faces_sr/sr16_noiseless/y0_\arabic{row}.jpg}
            &
        } \\ 
        & \forloop{row}{1}{\value{row} < 11}{ \includegraphics[width=\imwidth,height=\imwidth]{./figures/appendix_faces_sr/sr16_noiseless/\arabic{row}_s0_-1.jpg}
            &
        } \\ 
        \multicolumn{12}{c}{~} \\
        \multirow{2}{*}{\raisebox{0cm}[0pt][0pt]{\rotatebox[origin=c]{90}{$16\times$ noisy}}}
        & \forloop{row}{1}{\value{row} < 11}{ \includegraphics[width=\imwidth,height=\imwidth]{./figures/appendix_faces_sr/sr16_noisy/y0_\arabic{row}.jpg}
            &
        } \\ 
        & \forloop{row}{1}{\value{row} < 11}{ \includegraphics[width=\imwidth,height=\imwidth]{./figures/appendix_faces_sr/sr16_noisy/\arabic{row}_s0_-1.jpg}
            &
        }
    \end{tabular}
    \caption{Pairs of low-res and recovered $256\times256$ face images with a 20-step DDRM. Noisy low-res images contain noise with a standard deviation of $\sigma_{\rvy} = 0.1$.}
    \label{fig:apdx_faces_sr}
\end{figure}
\renewcommand{\imwidth}{1.6cm}
\begin{figure}
    \centering
    \def\arraystretch{0.2}
    \setlength\tabcolsep{0.005cm}
    \begin{tabular}{ccccccc}
    {Original} & ~~~ & \multicolumn{2}{c}{Inpainting} & ~~~ & \multicolumn{2}{c}{Deblurring} \\
    \forloop{row}{1}{\value{row} < 11}{
        \includegraphics[width=\imwidth,height=\imwidth]{./figures/appendix_faces_other/inp_noisy/orig_\arabic{row}.jpg}
        & ~ &
        \includegraphics[width=\imwidth,height=\imwidth]{./figures/appendix_faces_other/inp_noisy/y0_\arabic{row}.jpg}
        &
        \includegraphics[width=\imwidth,height=\imwidth]{./figures/appendix_faces_other/inp_noisy/\arabic{row}_s0_-1.jpg}
        & ~ &
        \includegraphics[width=\imwidth,height=\imwidth]{./figures/appendix_faces_other/deblur_noisy/y0_\arabic{row}.jpg}
        &
        \includegraphics[width=\imwidth,height=\imwidth]{./figures/appendix_faces_other/deblur_noisy/\arabic{row}_s0_-1.jpg}
        \\
    }
    \end{tabular}
    \caption{Pairs of degraded and recovered $256\times256$ face images with a 20-step DDRM. Degraded images contain noise with a standard deviation of $\sigma_{\rvy} = 0.1$.} 
    \label{fig:appendix_faces_other}
\end{figure}
\renewcommand{\imwidth}{1.6cm}
\newcommand{\imlimit}{4}
\begin{figure}
    \centering
    \def\arraystretch{0.2}
    \setlength\tabcolsep{0.005cm}
    \begin{tabular}{l cc cccc cc}
        \multirow{2}{*}{\raisebox{0cm}[0pt][0pt]{\rotatebox[origin=c]{90}{$4\times$ super-res}}}
        & \includegraphics[width=\imwidth,height=\imwidth]{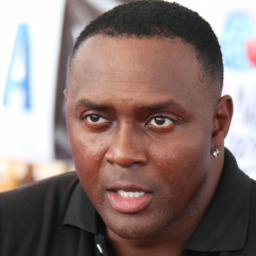} 
        & \includegraphics[width=\imwidth,height=\imwidth]{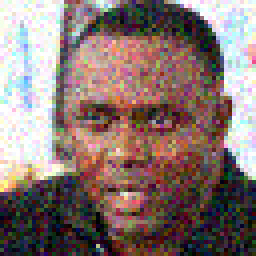}
        & \forloop{row}{0}{\value{row} < \imlimit}{
            \includegraphics[width=\imwidth,height=\imwidth]{./figures/appendix_diversity/sr4/0_s\arabic{row}_-1.jpg}
            &
        } 
        \includegraphics[width=\imwidth,height=\imwidth]{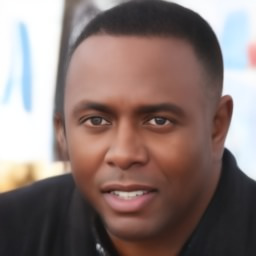}
        & \includegraphics[width=\imwidth,height=\imwidth]{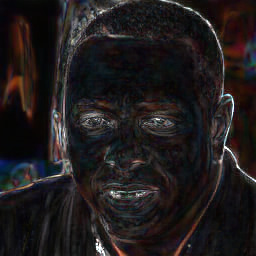}
        \\ 
        & \includegraphics[width=\imwidth,height=\imwidth]{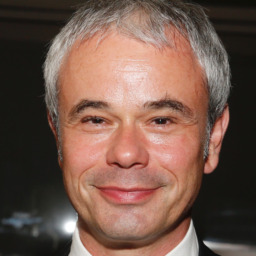}
        & \includegraphics[width=\imwidth,height=\imwidth]{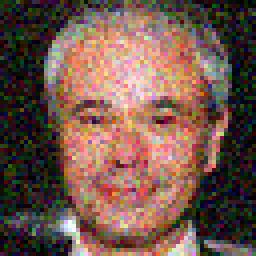}
        & \forloop{row}{0}{\value{row} < \imlimit}{
            \includegraphics[width=\imwidth,height=\imwidth]{./figures/appendix_diversity/sr4/2_s\arabic{row}_-1.jpg}
            &
        } 
        \includegraphics[width=\imwidth,height=\imwidth]{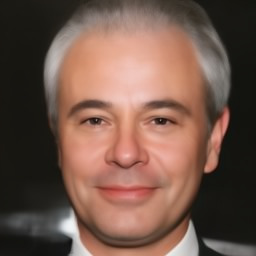}
        & \includegraphics[width=\imwidth,height=\imwidth]{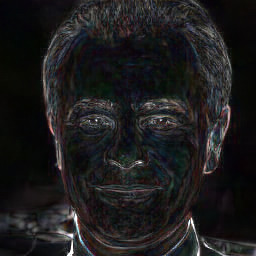}
        \\ 
        \multicolumn{9}{c}{~} \\
        
        \multirow{2}{*}{\raisebox{0cm}[0pt][0pt]{\rotatebox[origin=c]{90}{$8\times$ super-res}}}
        & \includegraphics[width=\imwidth,height=\imwidth]{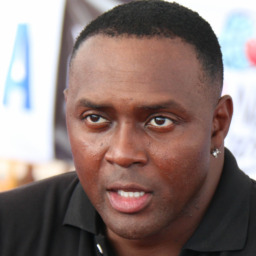}
        & \includegraphics[width=\imwidth,height=\imwidth]{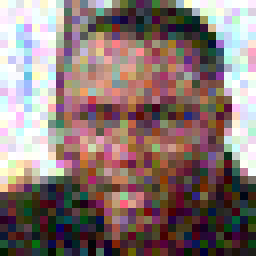}
        & \forloop{row}{0}{\value{row} < \imlimit}{
            \includegraphics[width=\imwidth,height=\imwidth]{./figures/appendix_diversity/sr8/0_s\arabic{row}_-1.jpg}
            &
        } 
        \includegraphics[width=\imwidth,height=\imwidth]{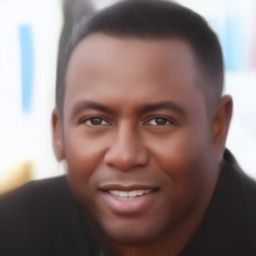}
        & \includegraphics[width=\imwidth,height=\imwidth]{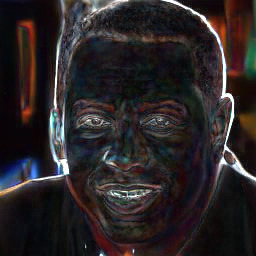}
        \\ 
        & \includegraphics[width=\imwidth,height=\imwidth]{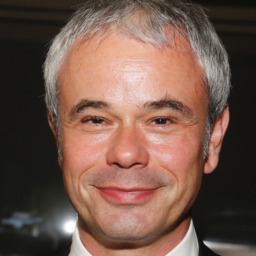}
        & \includegraphics[width=\imwidth,height=\imwidth]{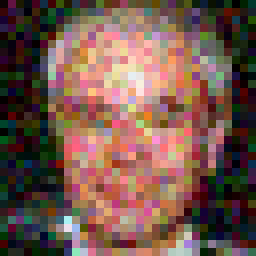}
        & \forloop{row}{0}{\value{row} < \imlimit}{
            \includegraphics[width=\imwidth,height=\imwidth]{./figures/appendix_diversity/sr8/2_s\arabic{row}_-1.jpg}
            &
        } 
        \includegraphics[width=\imwidth,height=\imwidth]{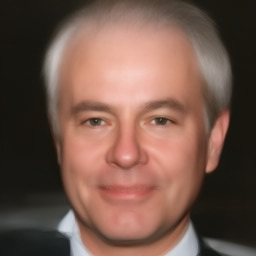}
        & \includegraphics[width=\imwidth,height=\imwidth]{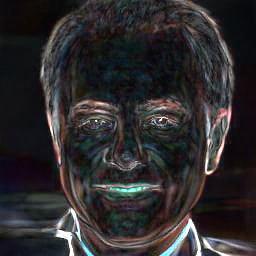}
        \\ 
        \multicolumn{9}{c}{~} \\
        
        \multirow{2}{*}{\raisebox{0cm}[0pt][0pt]{\rotatebox[origin=c]{90}{$16\times$ super-res}}}
        & \includegraphics[width=\imwidth,height=\imwidth]{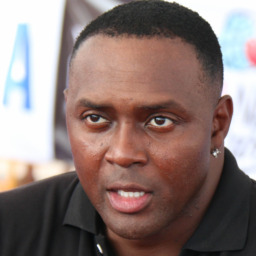}
        & \includegraphics[width=\imwidth,height=\imwidth]{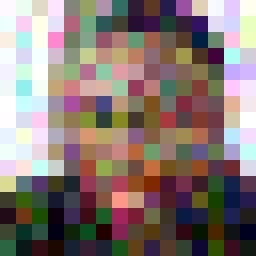}
        & \forloop{row}{0}{\value{row} < \imlimit}{
            \includegraphics[width=\imwidth,height=\imwidth]{./figures/appendix_diversity/sr16/0_s\arabic{row}_-1.jpg}
            &
        } 
        \includegraphics[width=\imwidth,height=\imwidth]{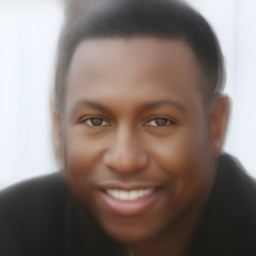}
        & \includegraphics[width=\imwidth,height=\imwidth]{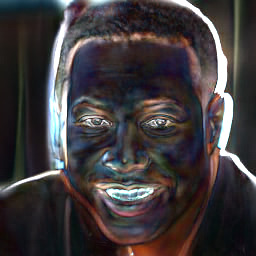}
        \\ 
        & \includegraphics[width=\imwidth,height=\imwidth]{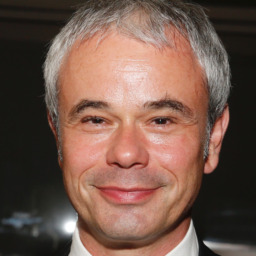}
        & \includegraphics[width=\imwidth,height=\imwidth]{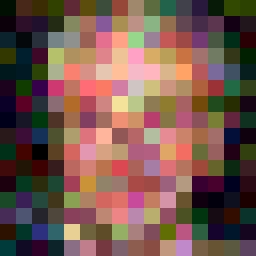}
        & \forloop{row}{0}{\value{row} < \imlimit}{
            \includegraphics[width=\imwidth,height=\imwidth]{./figures/appendix_diversity/sr16/2_s\arabic{row}_-1.jpg}
            &
        } 
        \includegraphics[width=\imwidth,height=\imwidth]{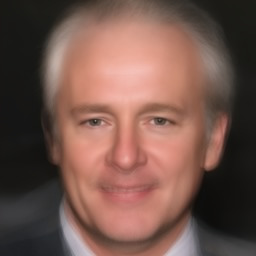}
        & \includegraphics[width=\imwidth,height=\imwidth]{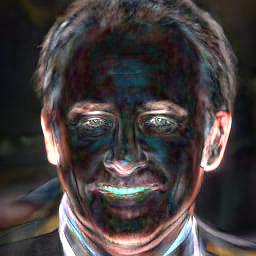}
        \\ 
        \multicolumn{9}{c}{~} \\
        
        \multirow{2}{*}{\raisebox{0cm}[0pt][0pt]{\rotatebox[origin=c]{90}{Inpainting}}}
        & \includegraphics[width=\imwidth,height=\imwidth]{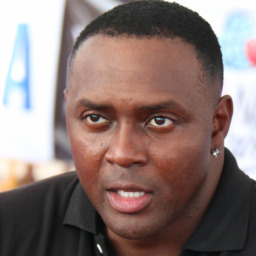}
        & \includegraphics[width=\imwidth,height=\imwidth]{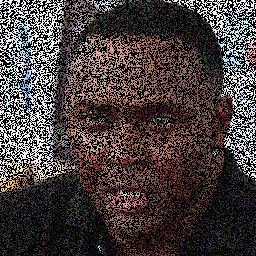}
        & \forloop{row}{0}{\value{row} < \imlimit}{
            \includegraphics[width=\imwidth,height=\imwidth]{./figures/appendix_diversity/inp/0_s\arabic{row}_-1.jpg}
            &
        } 
        \includegraphics[width=\imwidth,height=\imwidth]{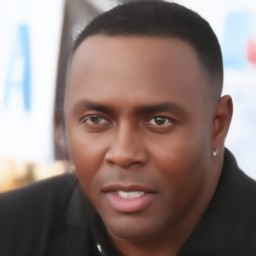}
        & \includegraphics[width=\imwidth,height=\imwidth]{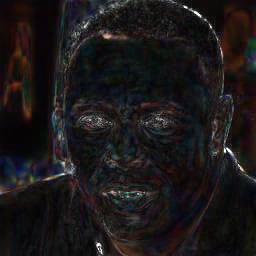}
        \\ 
        & \includegraphics[width=\imwidth,height=\imwidth]{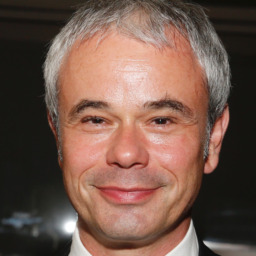}
        & \includegraphics[width=\imwidth,height=\imwidth]{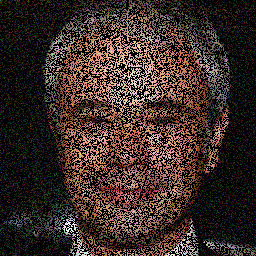}
        & \forloop{row}{0}{\value{row} < \imlimit}{
            \includegraphics[width=\imwidth,height=\imwidth]{./figures/appendix_diversity/inp/2_s\arabic{row}_-1.jpg}
            &
        } 
        \includegraphics[width=\imwidth,height=\imwidth]{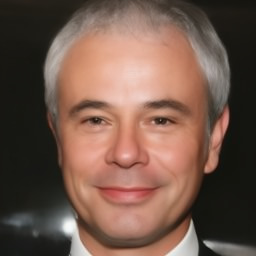}
        & \includegraphics[width=\imwidth,height=\imwidth]{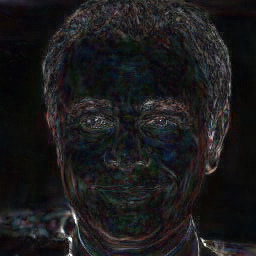}
        \\ 
        \multicolumn{9}{c}{~} \\
        
        \multirow{2}{*}{\raisebox{0cm}[0pt][0pt]{\rotatebox[origin=c]{90}{Deblurring}}}
        & \includegraphics[width=\imwidth,height=\imwidth]{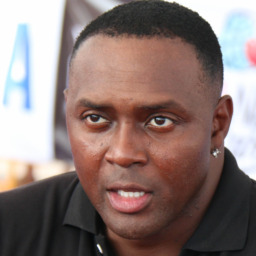}
        & \includegraphics[width=\imwidth,height=\imwidth]{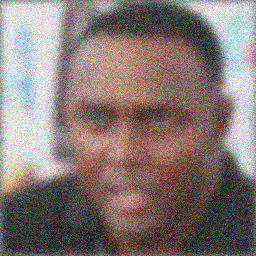}
        & \forloop{row}{0}{\value{row} < \imlimit}{
            \includegraphics[width=\imwidth,height=\imwidth]{./figures/appendix_diversity/deblur/0_s\arabic{row}_-1.jpg}
            &
        } 
        \includegraphics[width=\imwidth,height=\imwidth]{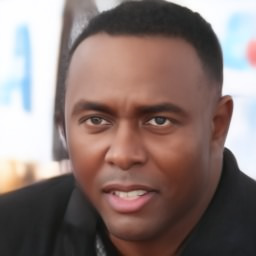}
        & \includegraphics[width=\imwidth,height=\imwidth]{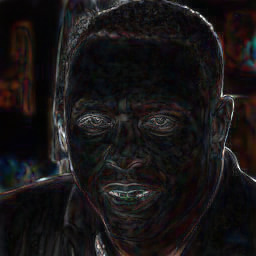}
        \\ 
        & \includegraphics[width=\imwidth,height=\imwidth]{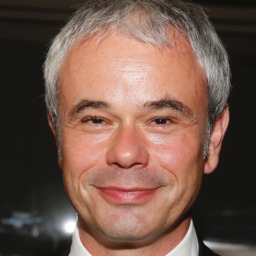}
        & \includegraphics[width=\imwidth,height=\imwidth]{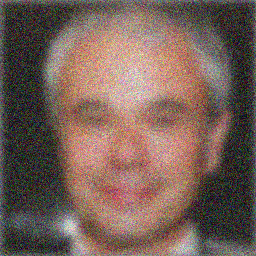}
        & \forloop{row}{0}{\value{row} < \imlimit}{
            \includegraphics[width=\imwidth,height=\imwidth]{./figures/appendix_diversity/deblur/2_s\arabic{row}_-1.jpg}
            &
        } 
        \includegraphics[width=\imwidth,height=\imwidth]{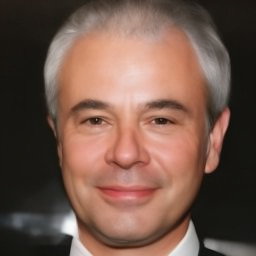}
        & \includegraphics[width=\imwidth,height=\imwidth]{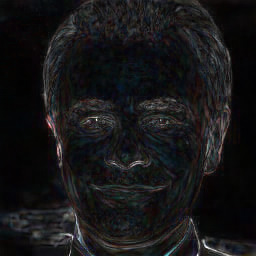}
        \\ 
        \multicolumn{9}{c}{~} \\
        & Original & Degraded & \multicolumn{\imlimit}{c}{Samples from a $20$-step DDRM} & Mean & std
    \end{tabular}
    \caption{Original, degraded, and $6$ recovered $256\times256$ face images with a 20-step DDRM. Degraded images contain noise with a standard deviation of $\sigma_{\rvy} = 0.1$. The mean and standard deviation (scaled by $4$) of the sampled solution is shown.}
    \label{fig:apdx_div}
\end{figure}
\renewcommand{\imwidth}{2.1cm}
\begin{figure}
    \centering
    \def\arraystretch{0.2}
    \setlength\tabcolsep{0.005cm}
    \begin{tabular}{l cccccc l}
    \raisebox{0.9cm}[0pt][0pt]{\rotatebox[origin=c]{90}{Original}} &
        \forloop{row}{0}{\value{row} < 6}{
            \includegraphics[width=\imwidth,height=\imwidth]{./figures/appendix_bedroom/inp/orig_\arabic{row}.jpg}
            &
        } \\
        \multicolumn{8}{c}{~} \\
        \multirow{2}{*}{\raisebox{0cm}[0pt][0pt]{\rotatebox[origin=c]{90}{Inpainting}}}
        & \forloop{row}{0}{\value{row} < 6}{
            \includegraphics[width=\imwidth,height=\imwidth]{./figures/appendix_bedroom/inp/y0_\arabic{row}.jpg}
            &
        } \\ 
        & \forloop{row}{0}{\value{row} < 6}{
            \includegraphics[width=\imwidth,height=\imwidth]{./figures/appendix_bedroom/inp/\arabic{row}_s0_-1.jpg}
            &
        } \\ 
        \multicolumn{8}{c}{~} \\
        \multirow{2}{*}{\raisebox{0cm}[0pt][0pt]{\rotatebox[origin=c]{90}{Colorization}}}
        & \forloop{row}{0}{\value{row} < 6}{
            \includegraphics[width=\imwidth,height=\imwidth]{./figures/appendix_bedroom/color/y0_\arabic{row}.jpg}
            &
        } \\ 
        & \forloop{row}{0}{\value{row} < 6}{
            \includegraphics[width=\imwidth,height=\imwidth]{./figures/appendix_bedroom/color/\arabic{row}_s0_-1.jpg}
            &
        } \\ 
        \multicolumn{8}{c}{~} \\
        \multirow{2}{*}{\raisebox{0cm}[0pt][0pt]{\rotatebox[origin=c]{90}{Deblurring}}}
        & \forloop{row}{0}{\value{row} < 6}{
            \includegraphics[width=\imwidth,height=\imwidth]{./figures/appendix_bedroom/deblur/y0_\arabic{row}.jpg}
            &
        } \\ 
        & \forloop{row}{0}{\value{row} < 6}{
            \includegraphics[width=\imwidth,height=\imwidth]{./figures/appendix_bedroom/deblur/\arabic{row}_s0_-1.jpg}
            &
        } \\ 
        \multicolumn{8}{c}{~} \\
        \multirow{2}{*}{\raisebox{0cm}[0pt][0pt]{\rotatebox[origin=c]{90}{$4\times$ super-res}}}
        & \forloop{row}{0}{\value{row} < 6}{
            \includegraphics[width=\imwidth,height=\imwidth]{./figures/appendix_bedroom/sr4/y0_\arabic{row}.jpg}
            &
        } \\ 
        & \forloop{row}{0}{\value{row} < 6}{
            \includegraphics[width=\imwidth,height=\imwidth]{./figures/appendix_bedroom/sr4/\arabic{row}_s0_-1.jpg}
            &
        } \\ 
    \end{tabular}
    \caption{Pairs of degraded and recovered $256\times256$ bedroom images with a 20-step DDRM. Degraded images contain noise with a standard deviation of $\sigma_{\rvy} = 0.05$.}
    \label{fig:apdx_bedroom}
\end{figure}
\renewcommand{\imwidth}{1.5cm}
\renewcommand{\imlimit}{9}
\begin{figure}
    \centering
    \def\arraystretch{0.2}
    \setlength\tabcolsep{0.005cm}
    \begin{tabular}{l ccccccccc l}
    \raisebox{0.7cm}[0pt][0pt]{\rotatebox[origin=c]{90}{Original}} &
        \forloop{row}{0}{\value{row} < \imlimit}{
            \includegraphics[width=\imwidth,height=\imwidth]{./figures/appendix_cat/inp/orig_\arabic{row}.jpg}
            &
        } \\
        \multicolumn{11}{c}{~} \\
        \multirow{2}{*}{\raisebox{0cm}[0pt][0pt]{\rotatebox[origin=c]{90}{Inpainting}}}
        & \forloop{row}{0}{\value{row} < \imlimit}{
            \includegraphics[width=\imwidth,height=\imwidth]{./figures/appendix_cat/inp/y0_\arabic{row}.jpg}
            &
        } \\ 
        & \forloop{row}{0}{\value{row} < \imlimit}{
            \includegraphics[width=\imwidth,height=\imwidth]{./figures/appendix_cat/inp/\arabic{row}_s0_-1.jpg}
            &
        } \\ 
        \multicolumn{11}{c}{~} \\
        \multirow{2}{*}{\raisebox{0cm}[0pt][0pt]{\rotatebox[origin=c]{90}{Colorization}}}
        & \forloop{row}{0}{\value{row} < \imlimit}{
            \includegraphics[width=\imwidth,height=\imwidth]{./figures/appendix_cat/color/y0_\arabic{row}.jpg}
            &
        } \\ 
        & \forloop{row}{0}{\value{row} < \imlimit}{
            \includegraphics[width=\imwidth,height=\imwidth]{./figures/appendix_cat/color/\arabic{row}_s0_-1.jpg}
            &
        } \\ 
        \multicolumn{11}{c}{~} \\
        \multirow{2}{*}{\raisebox{0cm}[0pt][0pt]{\rotatebox[origin=c]{90}{Deblurring}}}
        & \forloop{row}{0}{\value{row} < \imlimit}{
            \includegraphics[width=\imwidth,height=\imwidth]{./figures/appendix_cat/deblur/y0_\arabic{row}.jpg}
            &
        } \\ 
        & \forloop{row}{0}{\value{row} < \imlimit}{
            \includegraphics[width=\imwidth,height=\imwidth]{./figures/appendix_cat/deblur/\arabic{row}_s0_-1.jpg}
            &
        } \\ 
        \multicolumn{11}{c}{~} \\
        \multirow{2}{*}{\raisebox{0cm}[0pt][0pt]{\rotatebox[origin=c]{90}{$4\times$ super-res}}}
        & \forloop{row}{0}{\value{row} < \imlimit}{
            \includegraphics[width=\imwidth,height=\imwidth]{./figures/appendix_cat/sr4/y0_\arabic{row}.jpg}
            &
        } \\ 
        & \forloop{row}{0}{\value{row} < \imlimit}{
            \includegraphics[width=\imwidth,height=\imwidth]{./figures/appendix_cat/sr4/\arabic{row}_s0_-1.jpg}
            &
        } \\ 
    \end{tabular}
    \caption{Pairs of degraded and recovered $256\times256$ cat images with a 20-step DDRM. Degraded images contain noise with a standard deviation of $\sigma_{\rvy} = 0.05$.}
    \label{fig:apdx_cat}
\end{figure}
\renewcommand{\imwidth}{2.1cm}
\begin{figure}
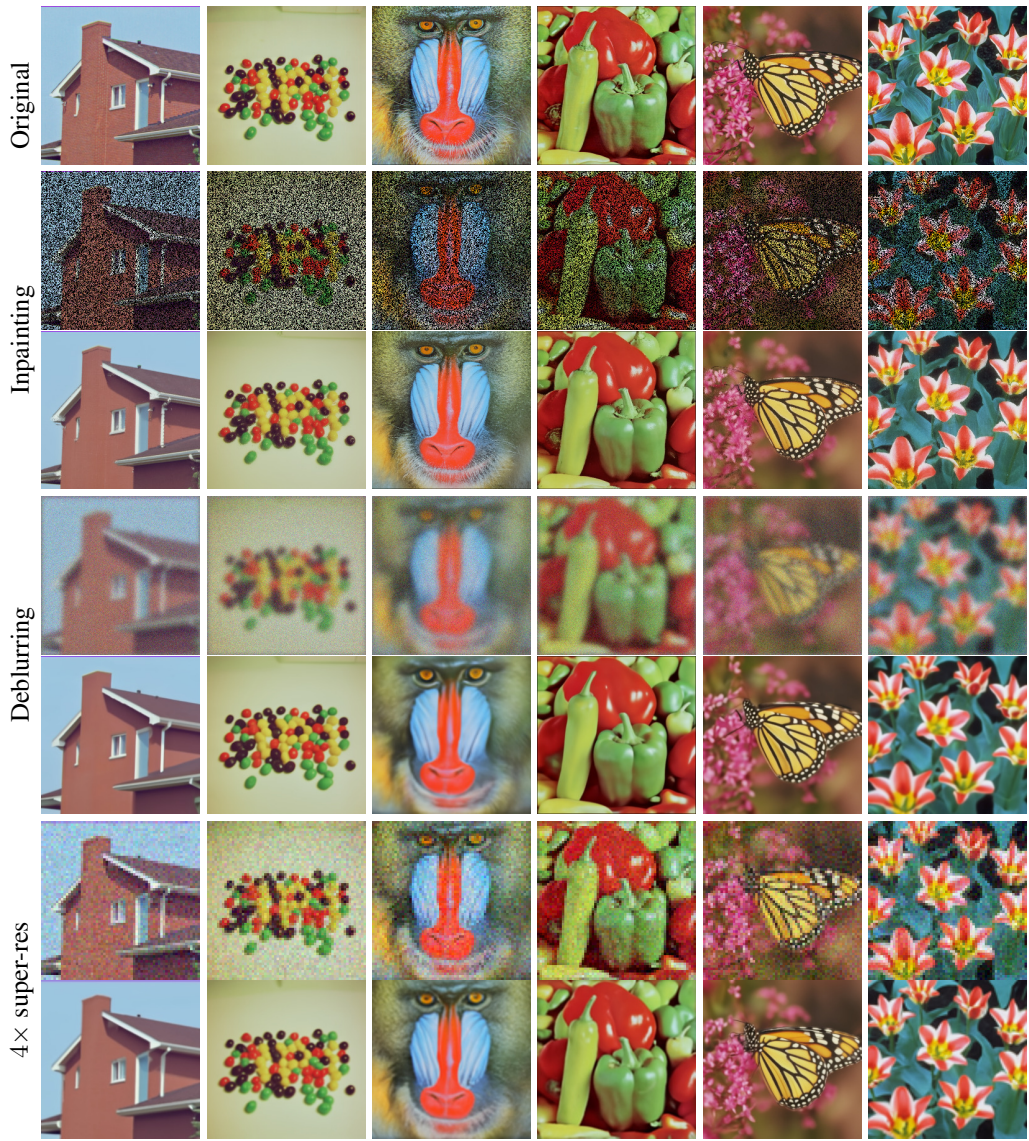

    \centering
    \def\arraystretch{0.2}
    \setlength\tabcolsep{0.005cm}
    \begin{tabular}{l cccccc l}
    \raisebox{0.7cm}[0pt][0pt]{\rotatebox[origin=c]{90}{Original}} &
        \forloop{row}{0}{\value{row} < 6}{
            \includegraphics[width=\imwidth,height=\imwidth]{./figures/appendix_ood/sr4_noisy/orig_\arabic{row}.jpg}
            &
        } \\
        \multicolumn{8}{c}{~} \\
        \multirow{2}{*}{\raisebox{0cm}[0pt][0pt]{\rotatebox[origin=c]{90}{Inpainting}}}
        & \forloop{row}{0}{\value{row} < 6}{
            \includegraphics[width=\imwidth,height=\imwidth]{./figures/appendix_ood/inp_noisy/y0_\arabic{row}.jpg}
            &
        } \\ 
        & \forloop{row}{0}{\value{row} < 6}{
            \includegraphics[width=\imwidth,height=\imwidth]{./figures/appendix_ood/inp_noisy/\arabic{row}_s0_-1.jpg}
            &
        } \\ 
        \multicolumn{8}{c}{~} \\
        \multirow{2}{*}{\raisebox{0cm}[0pt][0pt]{\rotatebox[origin=c]{90}{Deblurring}}}
        & \forloop{row}{0}{\value{row} < 6}{
            \includegraphics[width=\imwidth,height=\imwidth]{./figures/appendix_ood/deblur_noisy/y0_\arabic{row}.jpg}
            &
        } \\ 
        & \forloop{row}{0}{\value{row} < 6}{
            \includegraphics[width=\imwidth,height=\imwidth]{./figures/appendix_ood/deblur_noisy/\arabic{row}_s0_-1.jpg}
            &
        } \\ 
        \multicolumn{8}{c}{~} \\
        \multirow{2}{*}{\raisebox{0cm}[0pt][0pt]{\rotatebox[origin=c]{90}{$4\times$ super-res}}}
        & \forloop{row}{0}{\value{row} < 6}{
            \includegraphics[width=\imwidth,height=\imwidth]{./figures/appendix_ood/sr4_noisy/y0_\arabic{row}.jpg}
            &
        } \\ 
        & \forloop{row}{0}{\value{row} < 6}{
            \includegraphics[width=\imwidth,height=\imwidth]{./figures/appendix_ood/sr4_noisy/\arabic{row}_s0_-1.jpg}
            &
        } \\ 
    \end{tabular}
    \caption{Pairs of degraded and recovered $256\times256$ USC-SIPI images with a 20-step DDRM using an ImageNet diffusion model. Degraded images contain noise with a standard deviation of $\sigma_{\rvy} = 0.05$.}
    \label{fig:apdx_ood}
\end{figure}

\renewcommand{\imwidth}{1.5cm}
\begin{figure}
    \centering
    \def\arraystretch{0.5}
    \setlength\tabcolsep{0.005cm}
    \begin{tabular}{ccc c ccc c ccc}
        \forloop{row}{1}{\value{row} < 10}{
            \includegraphics[width=\imwidth,height=\imwidth]{./figures/appendix_in1k_sr4_noisy/ddrm20/orig_\arabic{row}00.jpg} &
            \includegraphics[width=\imwidth,height=\imwidth]{./figures/appendix_in1k_sr4_noisy/ddrm20/y0_\arabic{row}00.jpg} &
            \includegraphics[width=\imwidth,height=\imwidth]{./figures/appendix_in1k_sr4_noisy/ddrm20/\arabic{row}00_s0_-1.jpg} &
            ~ &
            \includegraphics[width=\imwidth,height=\imwidth]{./figures/appendix_in1k_sr4_noisy/ddrm20/orig_\arabic{row}25.jpg} &
            \includegraphics[width=\imwidth,height=\imwidth]{./figures/appendix_in1k_sr4_noisy/ddrm20/y0_\arabic{row}25.jpg} &
            \includegraphics[width=\imwidth,height=\imwidth]{./figures/appendix_in1k_sr4_noisy/ddrm20/\arabic{row}25_s0_-1.jpg}  &
            ~ &
            \includegraphics[width=\imwidth,height=\imwidth]{./figures/appendix_in1k_sr4_noisy/ddrm20/orig_\arabic{row}50.jpg} &
            \includegraphics[width=\imwidth,height=\imwidth]{./figures/appendix_in1k_sr4_noisy/ddrm20/y0_\arabic{row}50.jpg} &
            \includegraphics[width=\imwidth,height=\imwidth]{./figures/appendix_in1k_sr4_noisy/ddrm20/\arabic{row}50_s0_-1.jpg} \\
        }
        
        \includegraphics[width=\imwidth,height=\imwidth]{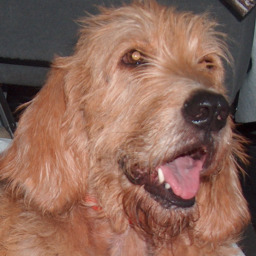} &
        \includegraphics[width=\imwidth,height=\imwidth]{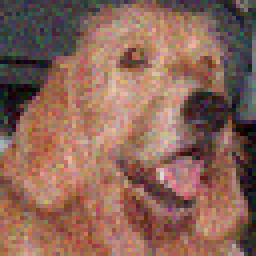} &
        \includegraphics[width=\imwidth,height=\imwidth]{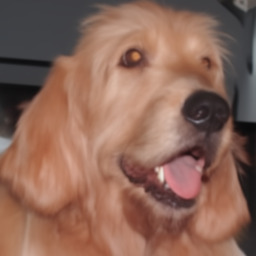} &
        ~ &
        \includegraphics[width=\imwidth,height=\imwidth]{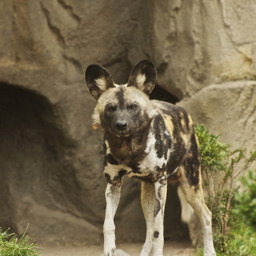} &
        \includegraphics[width=\imwidth,height=\imwidth]{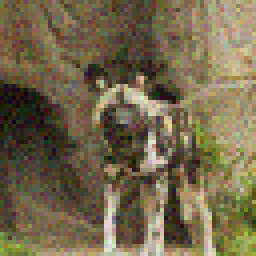} &
        \includegraphics[width=\imwidth,height=\imwidth]{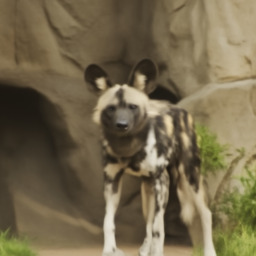} &
        ~ &
        \includegraphics[width=\imwidth,height=\imwidth]{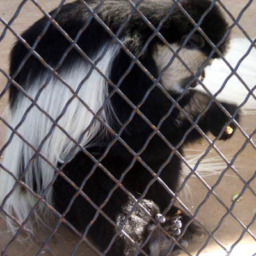} &
        \includegraphics[width=\imwidth,height=\imwidth]{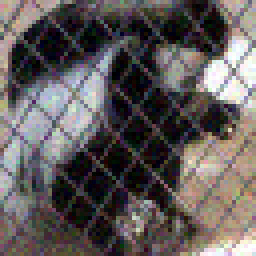} &
        \includegraphics[width=\imwidth,height=\imwidth]{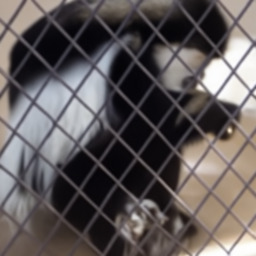} \\
        
        \includegraphics[width=\imwidth,height=\imwidth]{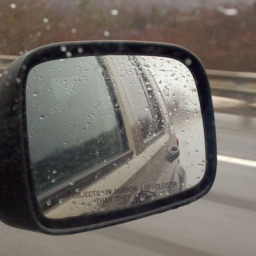} &
        \includegraphics[width=\imwidth,height=\imwidth]{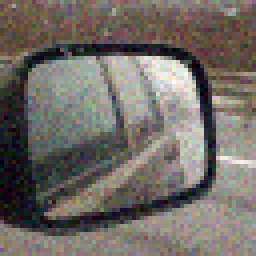} &
        \includegraphics[width=\imwidth,height=\imwidth]{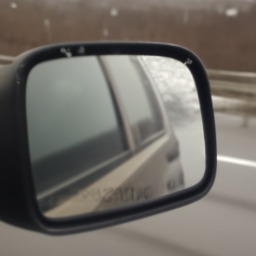} &
        ~ &
        \includegraphics[width=\imwidth,height=\imwidth]{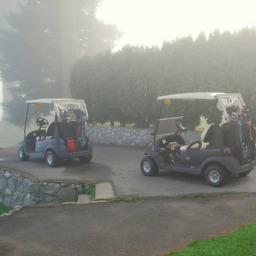} &
        \includegraphics[width=\imwidth,height=\imwidth]{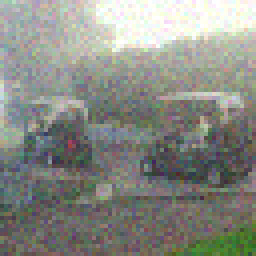} &
        \includegraphics[width=\imwidth,height=\imwidth]{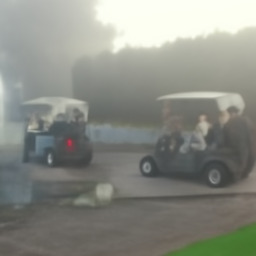} &
        ~ &
        \includegraphics[width=\imwidth,height=\imwidth]{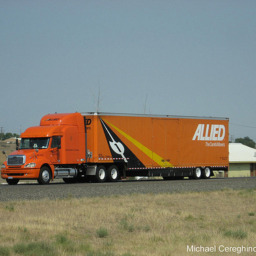} &
        \includegraphics[width=\imwidth,height=\imwidth]{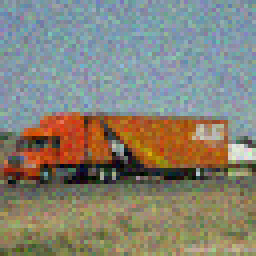} &
        \includegraphics[width=\imwidth,height=\imwidth]{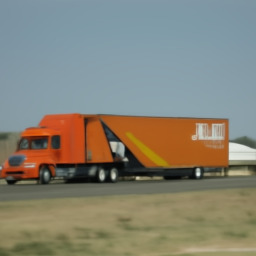} \\
        
        \includegraphics[width=\imwidth,height=\imwidth]{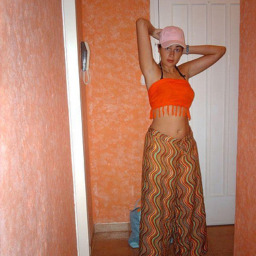} &
        \includegraphics[width=\imwidth,height=\imwidth]{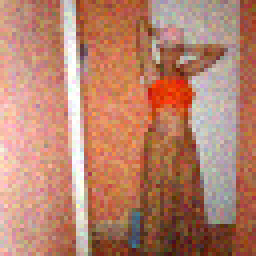} &
        \includegraphics[width=\imwidth,height=\imwidth]{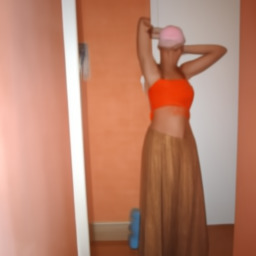} &
        ~ &
        \includegraphics[width=\imwidth,height=\imwidth]{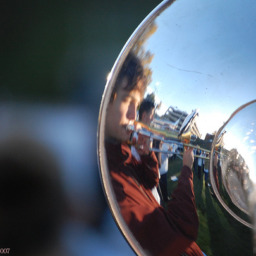} &
        \includegraphics[width=\imwidth,height=\imwidth]{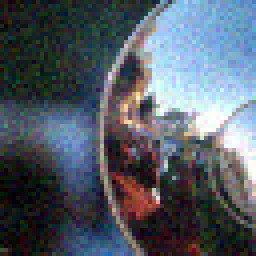} &
        \includegraphics[width=\imwidth,height=\imwidth]{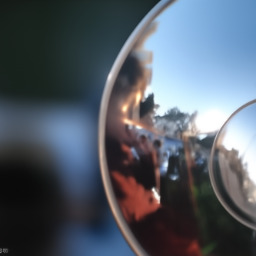} &
        ~ &
        \includegraphics[width=\imwidth,height=\imwidth]{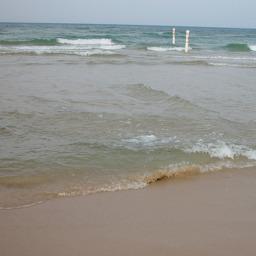} &
        \includegraphics[width=\imwidth,height=\imwidth]{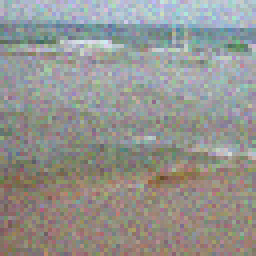} &
        \includegraphics[width=\imwidth,height=\imwidth]{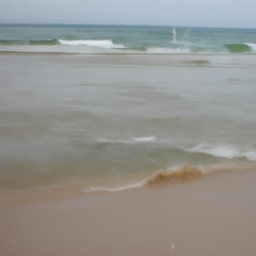} \\
    \end{tabular}
    \caption{Uncurated samples from the noisy $4\times$ super resolution ($\sigma_{\rvy} = 0.05$) task on $256\times256$ ImageNet 1K. Each triplet contains (from left to right): the original image, the low-res image, and the restored image with DDRM-$20$.}
    \label{fig:apdx_in1k_sr4}
\end{figure}
\renewcommand{\imwidth}{1.5cm}
\begin{figure}
    \centering
    \def\arraystretch{0.5}
    \setlength\tabcolsep{0.005cm}
    \begin{tabular}{ccc c ccc c ccc}
        \forloop{row}{1}{\value{row} < 10}{
            \includegraphics[width=\imwidth,height=\imwidth]{./figures/appendix_in1k_deblur_noisy/ddrm20/orig_\arabic{row}00.jpg} &
            \includegraphics[width=\imwidth,height=\imwidth]{./figures/appendix_in1k_deblur_noisy/ddrm20/y0_\arabic{row}00.jpg} &
            \includegraphics[width=\imwidth,height=\imwidth]{./figures/appendix_in1k_deblur_noisy/ddrm20/\arabic{row}00_s0_-1.jpg} &
            ~ &
            \includegraphics[width=\imwidth,height=\imwidth]{./figures/appendix_in1k_deblur_noisy/ddrm20/orig_\arabic{row}25.jpg} &
            \includegraphics[width=\imwidth,height=\imwidth]{./figures/appendix_in1k_deblur_noisy/ddrm20/y0_\arabic{row}25.jpg} &
            \includegraphics[width=\imwidth,height=\imwidth]{./figures/appendix_in1k_deblur_noisy/ddrm20/\arabic{row}25_s0_-1.jpg}  &
            ~ &
            \includegraphics[width=\imwidth,height=\imwidth]{./figures/appendix_in1k_deblur_noisy/ddrm20/orig_\arabic{row}50.jpg} &
            \includegraphics[width=\imwidth,height=\imwidth]{./figures/appendix_in1k_deblur_noisy/ddrm20/y0_\arabic{row}50.jpg} &
            \includegraphics[width=\imwidth,height=\imwidth]{./figures/appendix_in1k_deblur_noisy/ddrm20/\arabic{row}50_s0_-1.jpg} \\
        }
        
        \includegraphics[width=\imwidth,height=\imwidth]{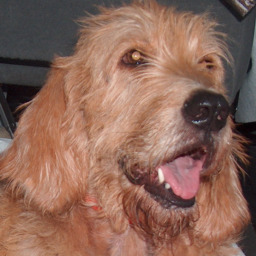} &
        \includegraphics[width=\imwidth,height=\imwidth]{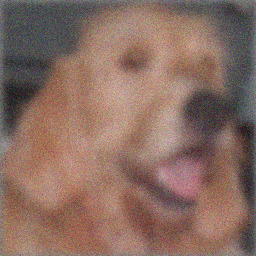} &
        \includegraphics[width=\imwidth,height=\imwidth]{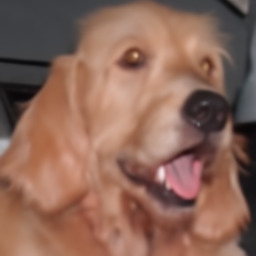} &
        ~ &
        \includegraphics[width=\imwidth,height=\imwidth]{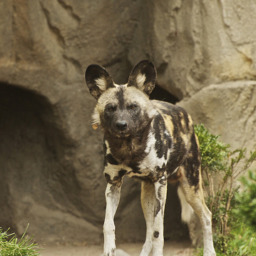} &
        \includegraphics[width=\imwidth,height=\imwidth]{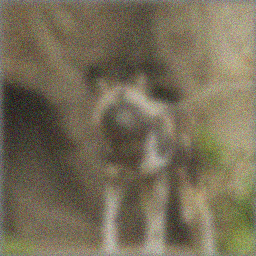} &
        \includegraphics[width=\imwidth,height=\imwidth]{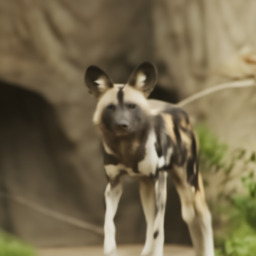} &
        ~ &
        \includegraphics[width=\imwidth,height=\imwidth]{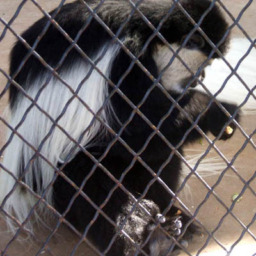} &
        \includegraphics[width=\imwidth,height=\imwidth]{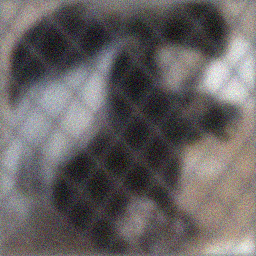} &
        \includegraphics[width=\imwidth,height=\imwidth]{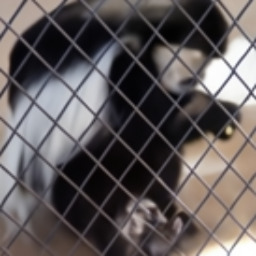} \\
        
        \includegraphics[width=\imwidth,height=\imwidth]{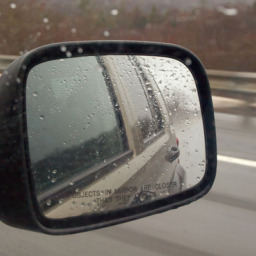} &
        \includegraphics[width=\imwidth,height=\imwidth]{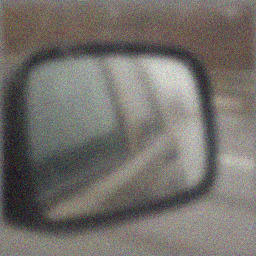} &
        \includegraphics[width=\imwidth,height=\imwidth]{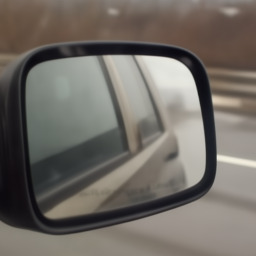} &
        ~ &
        \includegraphics[width=\imwidth,height=\imwidth]{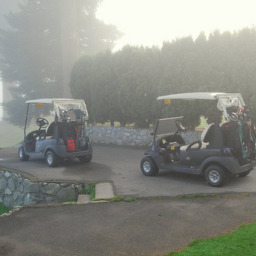} &
        \includegraphics[width=\imwidth,height=\imwidth]{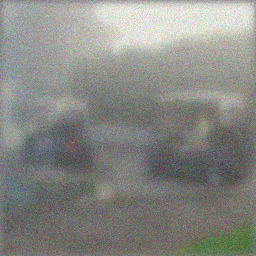} &
        \includegraphics[width=\imwidth,height=\imwidth]{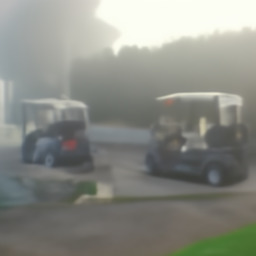} &
        ~ &
        \includegraphics[width=\imwidth,height=\imwidth]{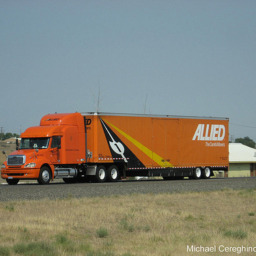} &
        \includegraphics[width=\imwidth,height=\imwidth]{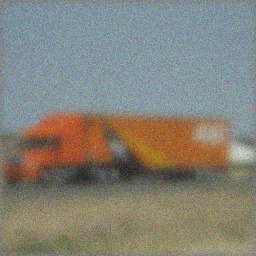} &
        \includegraphics[width=\imwidth,height=\imwidth]{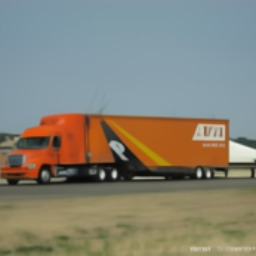} \\
        
        \includegraphics[width=\imwidth,height=\imwidth]{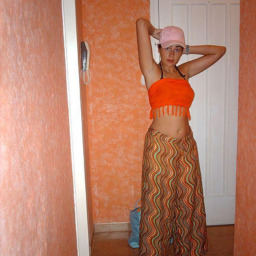} &
        \includegraphics[width=\imwidth,height=\imwidth]{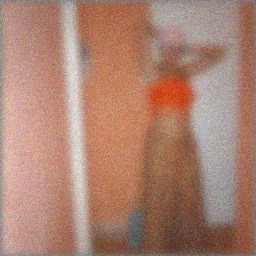} &
        \includegraphics[width=\imwidth,height=\imwidth]{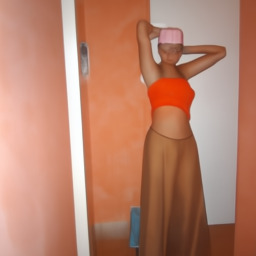} &
        ~ &
        \includegraphics[width=\imwidth,height=\imwidth]{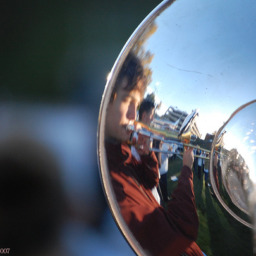} &
        \includegraphics[width=\imwidth,height=\imwidth]{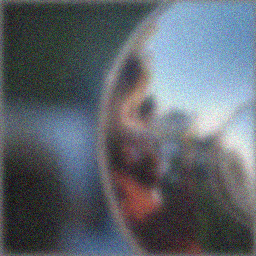} &
        \includegraphics[width=\imwidth,height=\imwidth]{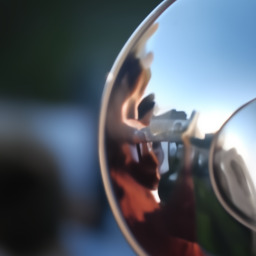} &
        ~ &
        \includegraphics[width=\imwidth,height=\imwidth]{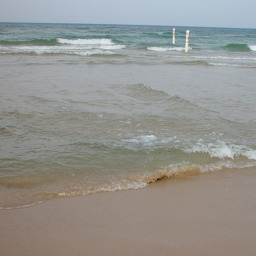} &
        \includegraphics[width=\imwidth,height=\imwidth]{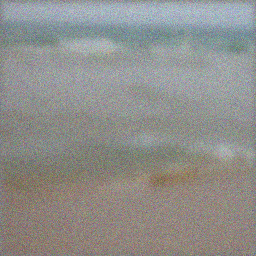} &
        \includegraphics[width=\imwidth,height=\imwidth]{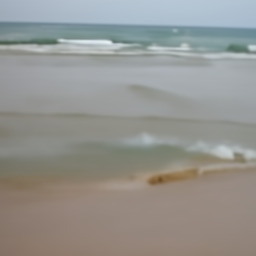} \\
    \end{tabular}
    \caption{Uncurated samples from the noisy deblurring ($\sigma_{\rvy} = 0.05$) task on $256\times256$ ImageNet 1K. Each triplet contains (from left to right): the original image, the blurry image, and the restored image with DDRM-$20$.}
    \label{fig:apdx_in1k_deblur}
\end{figure}

\begin{figure}[h]
    \centering
    \def\arraystretch{0.7}
    \setlength\tabcolsep{0.05cm}
    \begin{tabular}{cccc}
    \includegraphics[width=2.2cm,height=2.2cm]{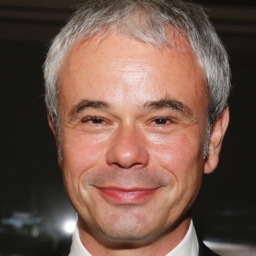} 
            & \includegraphics[width=2.2cm,height=2.2cm]{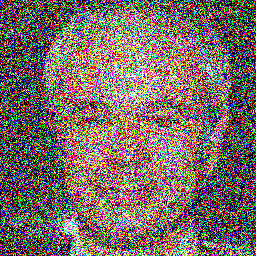} 
            & \includegraphics[width=2.2cm,height=2.2cm]{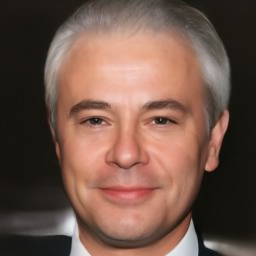}
            & \includegraphics[width=2.2cm,height=2.2cm]{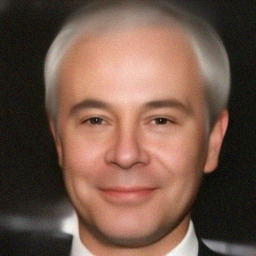} \\
        \includegraphics[width=2.2cm,height=2.2cm]{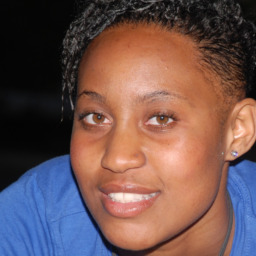} 
            & \includegraphics[width=2.2cm,height=2.2cm]{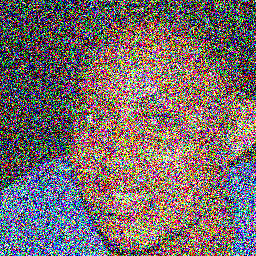} 
            & \includegraphics[width=2.2cm,height=2.2cm]{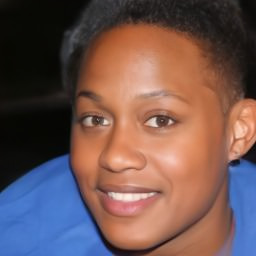}
            & \includegraphics[width=2.2cm,height=2.2cm]{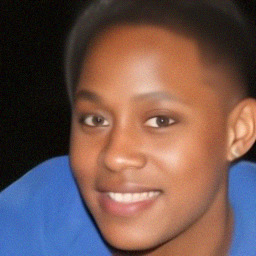} \\
        {Original} & {Noisy} & DDRM ($20$) & Denoised
    \end{tabular}
    \caption{Denoising ($\sigma_{\rvy} = 0.75$) face images. DDRM restores more fine details (\emph{e.g.} hair) than an MMSE denoiser. The denoiser used here is the denoising diffusion function $f_\theta(\rvx_{t}, t)$ used by DDRM, where $t$ minimizes $\left| \sigma_t - \sigma_\rvy \right|$.}
    \label{fig:faces_deno}
\end{figure}

\end{document}